\documentclass[10pt,letterpaper]{article}
\usepackage[top=0.85in,left=2.75in,footskip=0.75in]{geometry}

\newcommand{\ket}[1]{\vert #1 \rangle}

\newcommand{\braket}[2]{\langle #1 \vert #2 \rangle}
\usepackage{amsmath,amssymb}

\usepackage{changepage}

\usepackage[utf8x]{inputenc}

\usepackage{textcomp,marvosym}

\usepackage{cite}

\usepackage{nameref,hyperref}

\usepackage[right]{lineno}

\usepackage{microtype}
\DisableLigatures[f]{encoding = *, family = * }

\usepackage[table]{xcolor}

\usepackage{array}

\newcolumntype{+}{!{\vrule width 2pt}}

\newlength\savedwidth

\raggedright
\usepackage[aboveskip=1pt,labelfont=bf,labelsep=period,justification=raggedright,singlelinecheck=off]{caption}

\makeatletter
\renewcommand{\@biblabel}[1]{\quad#1.}
\makeatother

\usepackage{lastpage,fancyhdr,graphicx}
\usepackage{epstopdf}
\fancyheadoffset[L]{2.25in}
\fancyfootoffset[L]{2.25in}
\begin{document}
\vspace*{0.2in}

\begin{flushleft}
{\Large
\textbf\newline{Multiqubit and multilevel quantum reinforcement learning with quantum technologies} 
}
\newline
\\
F. A. C{\'a}rdenas-L{\'o}pez\textsuperscript{1,2,*},
L. Lamata\textsuperscript{3},
J. C. Retamal\textsuperscript{1,2},
E. Solano\textsuperscript{3,4,5}
\\
\bigskip
\textbf{1} Departamento de F\'isica, Universidad de Santiago de Chile (USACH), Santiago, Chile
\\
\textbf{2} Center for the Development of Nanoscience and Nanotechnology, Estaci\'on Central, Santiago, Chile
\\
\textbf{3} Department of Physical Chemistry, University of the Basque Country UPV/EHU, Bilbao, Spain
\\
\textbf{4} IKERBASQUE, Basque Foundation for Science, Bilbao, Spain
\\
\textbf{5} Department of Physics, Shanghai University, Shanghai, China
\\
\bigskip

%
%





Corresponding author\\
francisco.cardenas@usach.cl (FAC-L)
\end{flushleft}
\section*{Abstract}
We propose a protocol to perform quantum reinforcement learning with quantum technologies. At variance with recent results on quantum reinforcement learning with superconducting circuits, in our current protocol coherent feedback during the learning process is not required, enabling its implementation in a wide variety of quantum systems. We consider diverse possible scenarios for an agent, an environment, and a register that connects them, involving multiqubit and multilevel systems, as well as open-system dynamics. We finally propose possible implementations of this protocol in trapped ions and superconducting circuits. The field of quantum reinforcement learning with quantum technologies will enable enhanced quantum control, as well as more efficient machine learning calculations.


\section*{Introduction}
Machine Learning (ML) is a subfield of Artificial Intelligence (AI) that has attracted increasing attention in the last years. ML usually refers to {a computer program which can learn from experience E with respect to some class of task T and performance measure P, if its performance at tasks in T, as measured by P, improves with experience E }\cite{michalski2013machine}. In other words, Machine Learning addresses the problem of how a computer algorithm can be constructed to automatically improve with experience. Several applications in this field have been implemented such as handwriting pattern recognition \cite{Plamodon2000}, speech recognition \cite{Lee1989} and the development of a computer able to beat an expert Go player \cite{Nature.529.484}, just to name a few.\par
The learning process in ML can be divided in three types: supervised learning, unsupervised learning and reinforcement learning \cite{Russell}. In supervised machine learning, an initial data set has the function of training the system for later prediction making or to classify data. Usually, supervised learning problems are categorized into regression (continuous output) or classification (discrete output). Unsupervised learning allows one to address problems where the training data is not necessary and only correlations between subsets in the data (clustering) are considered and analyzed. Finally, reinforcement learning \cite{sutton1998reinforcement} differs from supervised and unsupervised learning in that it takes into account a scalar parameter (reward) to evaluate the input-output relation in a trial and error way. In this case, the system (so-called ``agent'') obtains information from its outer world (``environment'') to decide which is the better way to optimize itself, for adapting to the environment.

Quantum information processing (QIP) could contribute positively in the future in the development of the machine learning field, with several quantum algorithms for machine learning with significant possible gains with respect to their classical counterparts~\cite{wittek2014quantum,schuld2015introduction,adcock2015advances,biamonte2016quantum,DunjkoReview}. More specifically, quantum algorithms have been developed and in some cases implemented  for supervised and unsupervised learning problems \cite{bonner2003survey,Aimeur2013,lloyd2013quantum,PhysRevLett.113.130503,UnaiFeedback,support1,support2}. However, quantum reinforcement learning has not been widely explored and just a few results have been obtained up to now~\cite{RLDong,PhysRevX.4.031002,PhysRevLett.117.130501,crawford2016reinforcement,Lucas2017,Friis2015,Dunjko2015,Sriarunothai2017}. Related topics in biomimetic quantum technologies are quantum memristors~\cite{Mikel1,Mikel2,Mikel3,NoriMemristors}, as well as quantum Helmholtz and Boltzmann machines~\cite{Alejandro1,Alejandro2,Alejandro3}. These, together with quantum reinforcement learning, may set the stage for the future development of semi-autonomous quantum devices.

The field of quantum technologies has grown extensively in the past decade. In particular, two architectures which are very promising for the implementation of a quantum computer, in terms of numbers of qubits and gate fidelities, are trapped ions \cite{RevModPhys.75.281,HAFFNER2008155}   and superconducting circuits \cite{PhysRevA.75.032329,Clarke2008,wendin2016quantum}. Current technological progress in trapped ions has allowed us to implement quantum protocols with several ions involving high-fidelity single and two-qubit gates as well as high-fidelity readout~\cite{PhysRevLett.113.220501,PhysRevLett.117.060504}. Superconducting circuits have also proven to be an excellent platform to perform quantum information processing protocols because of their individual addressing and scalability. Two-qubit quantum gates have achieved fidelities larger than 99{\%} \cite{Barends2014,Barends2016} in this platform. Furthermore, technological progress in this architecture has made possible to build artificial atoms with high coherence time in coplanar \cite{PhysRevLett.111.080502} and 3D architecture \cite{PhysRevLett.107.240501}, allowing for the development of feedback control with superconducting circuits~\cite{PhysRevLett.109.240502,riste2015digital}. This feedback mechanism has  inspired protocols for quantum reinforcement learning with superconducting circuits~\cite{Lucas2017} where the feedback loop control allows one to reward and restart the system to obtain maximal learning fidelity.

Here, we propose a general protocol to perform quantum reinforcement learning with quantum technologies. We understand general in the sense that it goes beyond the context of qubits for embedding information in agent or environment. In this sense, and at variance with a previous result \cite{Lucas2017}, we extend the realm of the quantum reinforcement learning protocol to multi-qubit, multi-level, and open quantum systems, therefore permitting a wider set of scenarios. Our protocol considers a quantum system (the agent), which interacts with an external quantum  system (its environment) via an auxiliary quantum system (a register). The aim of our quantum reinforcement learning protocol is for the agent to acquire information from its environment and adapt to it, via a rewarding mechanism. In this fully quantum scenario the meaning of the learning process is the establishment of quantum correlations among the parties~\cite{PhysRevLett.117.130501}.  In our specific case, the quantum agent aims at attaining maximum quantum state overlap with the environment state, in the sense that local measurements on agent and environment will produce the same outcomes or, equivalently, that the agent and environment entangled final state is invariant under the exchange of these two subsystems. An interpretation of this outcome is that the agent can learn about the information embedded in the environment state, which has been consequently modified from a separable to an entangled state with the agent and registers. After this process we are in position of evaluating any figure of merit with the outcome measurements. Optimizing this figure of merit should be associated to a particular learning process probably requiring particular actions to be applied on the agent. Another possible result is obtained by considering projective measurements in the register systems. Only after these projective measurements agent and environment will be decoupled from them and the protocol assures that the former are in a pure correlated state, without needing to know any information about their initial states.  We analyze the case where the register subspace is larger than agent and environment subspaces. The inclusion of more elements in the register subspace allows for delaying the application of the rewarding criterion to the end of the quantum protocol. This fact will enable its implementation in a wider variety of quantum platforms, besides superconducting circuits with coherent feedback. We also study quantum reinforcement learning in the case where agent, environment and register are composed of qudits. In this case, we obtain that the maximal learning fidelity is achieved in a fixed number of steps in the qudit dimension, and this number scales polynomially with the number of subsystems in the environment subspace. In addition, we analyse quantum reinforcement learning in the situation where the environment is larger than the agent. We highlight two results: the first of them is obtained when considering that the register has the same elements than the environment. In this case, two rewarding criteria are needed to obtain maximal learning fidelity and the entanglement between the agent and a specific part of the environment is a key resource. The other case is the situation where the register has more elements than the environment. In this case, only one measurement is needed to obtain maximal learning fidelity and the environment-agent entanglement is not a key resource. Based on this fact, the rewarding criterion is applied at the end of the protocol. Finally, we describe how our quantum learning protocols can be implemented in quantum platforms as trapped ions and superconducting circuits
\section*{Quantum reinforcement learning protocol with final measurement}
Here, we introduce a protocol to perform quantum reinforcement learning, which introduces significant novelties with respect to the existing literature. Unlike a previous quantum reinforcement learning result~\cite{Lucas2017}, the protocol described here needs one measurement at the end of the procedure and no feedback, allowing for its implementation in a variety of quantum platforms including ions and photons. The improvement relies on adding more registers than before~\cite{Lucas2017} and making them interact conditionally with each other. The inclusion of ancillary systems has proven to be useful in several implementations of quantum information, because measurements on the ancillary system allow one in principle to obtain information about the main system without destroying it. Moreover, the measurement associated with the rewarding criterion is performed at the end of the protocol. This opens the possibility to implement quantum reinforcement learning protocols in architectures for which implementing coherent feedback may be a challenging problem.

\begin{figure}[!h]
	\centering
	\includegraphics[width=1\linewidth]{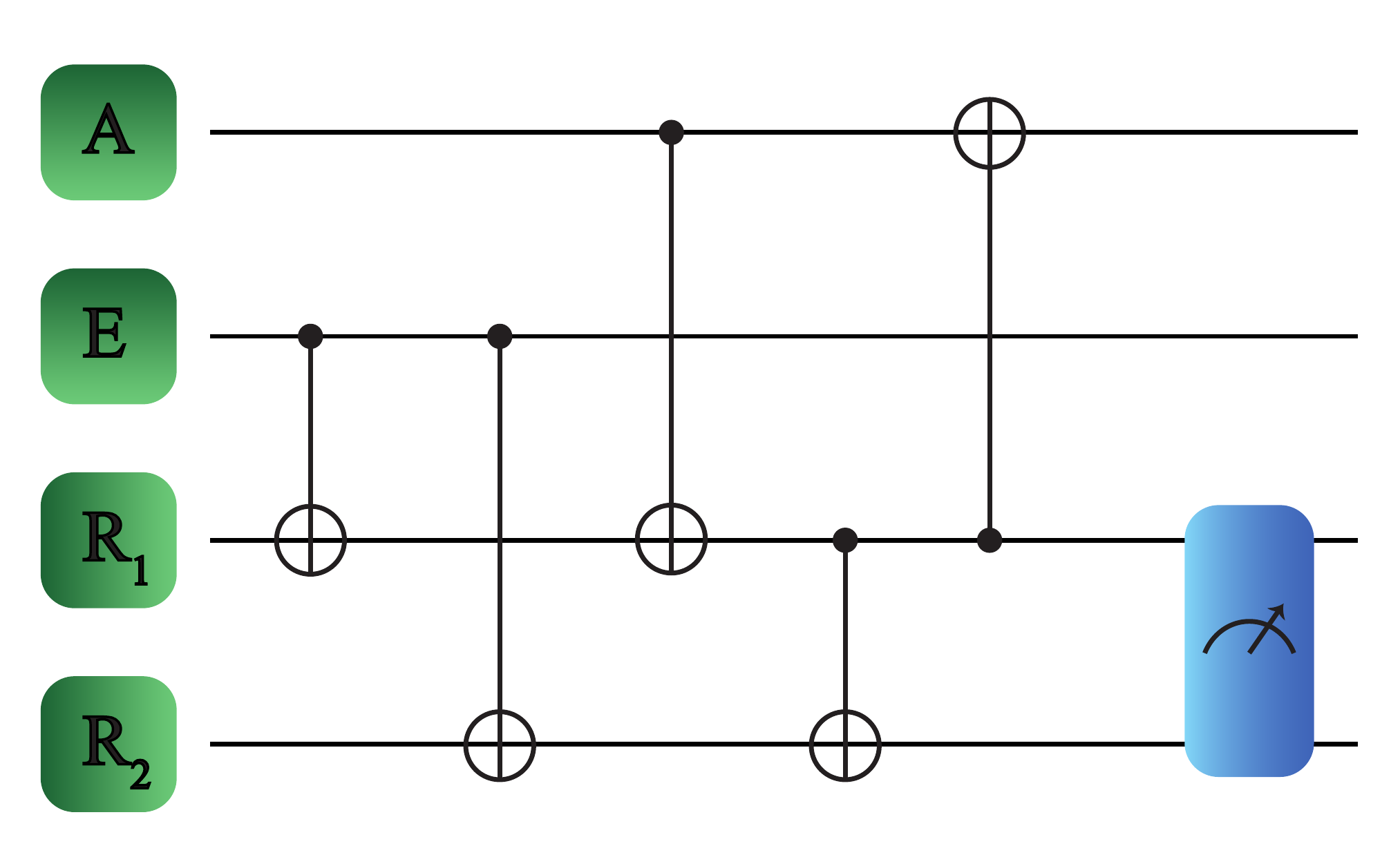}
	\caption{\textbf{Proposed protocol to perform quantum reinforcement learning with final measurement}. We consider a set composed of four qubits, corresponding to agent A, environment E, and registers $\rm{R}_1$ and $\rm{R}_2$. The considered interactions agent-register, register-register and environment-register consist of CNOT gates. The measurement in the register subspace is denoted by the rightmost box.}
	\label{Fig1}	
\end{figure}

The quantum reinforcement learning protocol described here works in the following way. We firstly consider an agent and environment, composed of one qubit each, and two register qubits, see \textbf{Fig. \ref{Fig1}}. The first step is to encode the environment information in the register states (usually this kind of operation in the context of classical reinforcement learning is called the action). Subsequently, the internal states of the registers interact conditionally with the agent (usually this kind of operation in classical reinforcement learning is called the percept). Finally, an agent-register interaction changes the agent state (partial rewarding mechanism). At this stage the rewarding criterion is satisfied, in the form of a correlated agent-environment state, in the sense that local measurements on agent and environment will produce the same outcomes. On the other hand, the agent-environment system is also entangled with the two registers, and in order to attain a correlated pure state of agent and environment, a single, final measurement may be performed on the two register states. This will produce an agent-environment state maximizing the learning fidelity defined as  $\mathcal{F}_{AE} = |\braket{\psi_A}{\phi_E}|$, where $\ket{\psi_A}$ is the agent state and $\ket{\phi_E}$ is the environment state, both after the protocol.
 
To perform our quantum reinforcement learning protocol we consider that initially agent and environment are in arbitrary single-qubit pure states, whereas the register states are in their ground state, namely

\begin{eqnarray}
\label{eqn1}
\{\ket{A} & = & \alpha_{A}^{0}\ket{0}_A + \alpha_{A}^{1}\ket{1}_A,\ket{E} =\alpha_{E}^{0}\ket{0}_E+\alpha_{E}^{1}\ket{1}_E,\ket{R}=\ket{0}_1\ket{0}_2 \}\\
\ket{\Psi}_0 & = & \ket{A}\ket{E}\ket{R}.
\end{eqnarray}
The first step in the protocol is to extract information from the environment, updating the information in the registers conditionally to the environment state. This process is done by applying a pair of CNOT gates in the environment-register subspace. Here, the first system is the control and the second the target,
\begin{eqnarray}
\label{eq2}
\ket{\Psi}_1 & = & U^{\rm{CNOT}}_{({E},{R_2})}U^{\rm{CNOT}}_{({E},{R_1})}\ket{\Psi}_0,\\
\ket{\Psi}_1 & = & \big(\alpha_{A}^{0}\ket{0}_A + \alpha_{A}^{1}\ket{1}_A\big)\big(\alpha_{E}^{0}\ket{0}_E\ket{0}_{1}\ket{0}_{2}+\alpha_{E}^{1}\ket{1}_E\ket{1}_{1}\ket{1}_{2}\big).
\end{eqnarray}
Then, the information encoded on the registers is updated conditional on the agent state. As the register subspace is larger than the agent subspace, we will choose which part of the register subspace will the agent update. Without loss of generality, let us assume that the register $R_1$ will be updated. The upgrade of agent subspace is performed by a CNOT gate acting in the $A-R_1$ subspace,  where the agent state is the control and the register is the target,
\begin{eqnarray}
\label{eq3}\nonumber
\ket{\Psi}_2 & = & U^{\rm{CNOT}}_{({A},{R_1})}\ket{\Psi}_1,\\\nonumber
\ket{\Psi}_2 & = & \big(\alpha_{A}^{0}\alpha_{E}^{0}\ket{0}_A\ket{0}_E\ket{0}_1\ket{0}_2 + \alpha_{A}^{0}\alpha_{E}^{1}\ket{0}_A\ket{1}_E\ket{1}_1\ket{1}_2+\alpha_{A}^{1}\alpha_{E}^{0}\ket{1}_A\ket{0}_E\ket{1}_1\ket{0}_2\\
&+& \alpha_{A}^{1}\alpha_{E}^{1}\ket{1}_A\ket{1}_E\ket{0}_1\ket{1}_2\big).
\end{eqnarray}
Subsequently, the register $R_2$ is also updated with respect to the $R_1$ state. This is accomplished by applying a CNOT gate in the register subspace, where $R_1$ acts as control and $R_2$ as target,
\begin{eqnarray}
\label{eq4}\nonumber
\ket{\Psi}_3  & = & U^{\rm{CNOT}}_{({R_1},{R_2})}\ket{\Psi}_2,\\\nonumber
\ket{\Psi}_3  & = & \big(\alpha_{A}^{0}\alpha_{E}^{0}\ket{0}_A\ket{0}_E\ket{0}_1\ket{0}_2 + \alpha_{A}^{0}\alpha_{E}^{1}\ket{0}_A\ket{1}_E\ket{1}_1\ket{0}_2 + \alpha_{A}^{1}\alpha_{E}^{0}\ket{1}_A\ket{0}_E\ket{1}_1\ket{1}_2\\
&+& \alpha_{A}^{1}\alpha_{E}^{1}\ket{1}_A\ket{1}_E\ket{0}_1\ket{1}_2\big).
\end{eqnarray}
Followingly, we update the agent state according to the information encoded in the register $R_1$.  This is done by applying a CNOT gate in the $R_1-A$ subspace, where $R_1$ is the control and $A$ is the target,
\begin{eqnarray}
\label{eq5}\nonumber
\ket{\Psi}_4  & = & U^{\rm{CNOT}}_{({R_1},{A})}\ket{\Psi}_3,\\\nonumber
\ket{\Psi}_4  & = & \big(\alpha_{A}^{0}\alpha_{E}^{0}\ket{0}_A\ket{0}_E\ket{0}_1\ket{0}_2 + \alpha_{A}^{0}\alpha_{E}^{1}\ket{1}_A\ket{1}_E\ket{1}_1\ket{0}_2 + \alpha_{A}^{1}\alpha_{E}^{0}\ket{0}_A\ket{0}_E\ket{1}_1\ket{1}_2\\
&+& \alpha_{A}^{1}\alpha_{E}^{1}\ket{1}_A\ket{1}_E\ket{0}_1\ket{1}_2\big).
\end{eqnarray}
We point out that, in the previous state, agent and environment are already maximally correlated, in the sense of having the same outcomes with respect to local measurements performed on either of them, or, equivalently, the state is invariant under particle exchange with respect to the agent-environment subsystem. We also remark that this state is general, valid for any initial agent and environment states. The fact that agent and environment get entangled with the two registers allows one to distinguish between identical agent-environment components that originate from different initial states, namely, to distinguish between states arising from $\alpha_{A}^{0}\alpha_{E}^{0}$ or $\alpha_{A}^{1}\alpha_{E}^{0}$, as well as from $\alpha_{A}^{0}\alpha_{E}^{1}$ or $\alpha_{A}^{1}\alpha_{E}^{1}$.\par

Finally, by performing a projective measurement on the register subspace, the rewarding criteron is satisfied. It is easy to show that, independently of the measurement outcome, the learning fidelity $\mathcal{F}_{AE} = |\braket{\psi_A}{\phi_E}|$  is maximal, given that agent and environment states end up being in the same state, either $|0\rangle$ or $|1\rangle$. In this case only one iteration of the protocol is sufficient in order that the agent adapts to the environment. Moreover, throughout the protocol, measurements on agent and/or environment are not required, which may allow its implementation in a variety of quantum platforms as trapped ions, superconducting circuits, and quantum photonics.\par

In our protocol, we do not need coherent feedback given that the registers entangle with agent and environment and as a result produce the desired agent-environment state that is invariant under permutation. It is true that the entanglement with the registers produces a mixed state in case the register states are discarded, but this is not a drawback in our protocol. Indeed, what our protocol does is, for arbitrary initial agent and environment states, which need not be known, to give a constructive way to produce a final agent-environment state perfectly correlated, in the sense of invariant under permutations in agent-environment subspace. This state is in general entangled, namely, quantum, and we do not need to perform any measurement on agent and environment during the protocol, namely, it can equally well work with photons, ions, and superconducting circuits, among others. After the production of the agent-environment-register entangled state, the registers are entangled with agent and environment, but this does not prevent us from measuring the registers at a certain desired time, and decoupling agent and environment from them. This way, we will not have measured agent and environment at any time of the protocol, and we can assure that they are perfectly correlated irrespective of their initial states, and without having any prior information about them. This may be useful, e.g., for distributing private keys in quantum cryptography for arbitrary, unknown, initial states, without the need to initialize agent and register in reference states.

\section*{Quantum reinforcement learning for multiqubit systems with final measurement}\label{Two-qubit_subspace}
\begin{figure}[!h]
\centering
\includegraphics[width=0.7\linewidth]{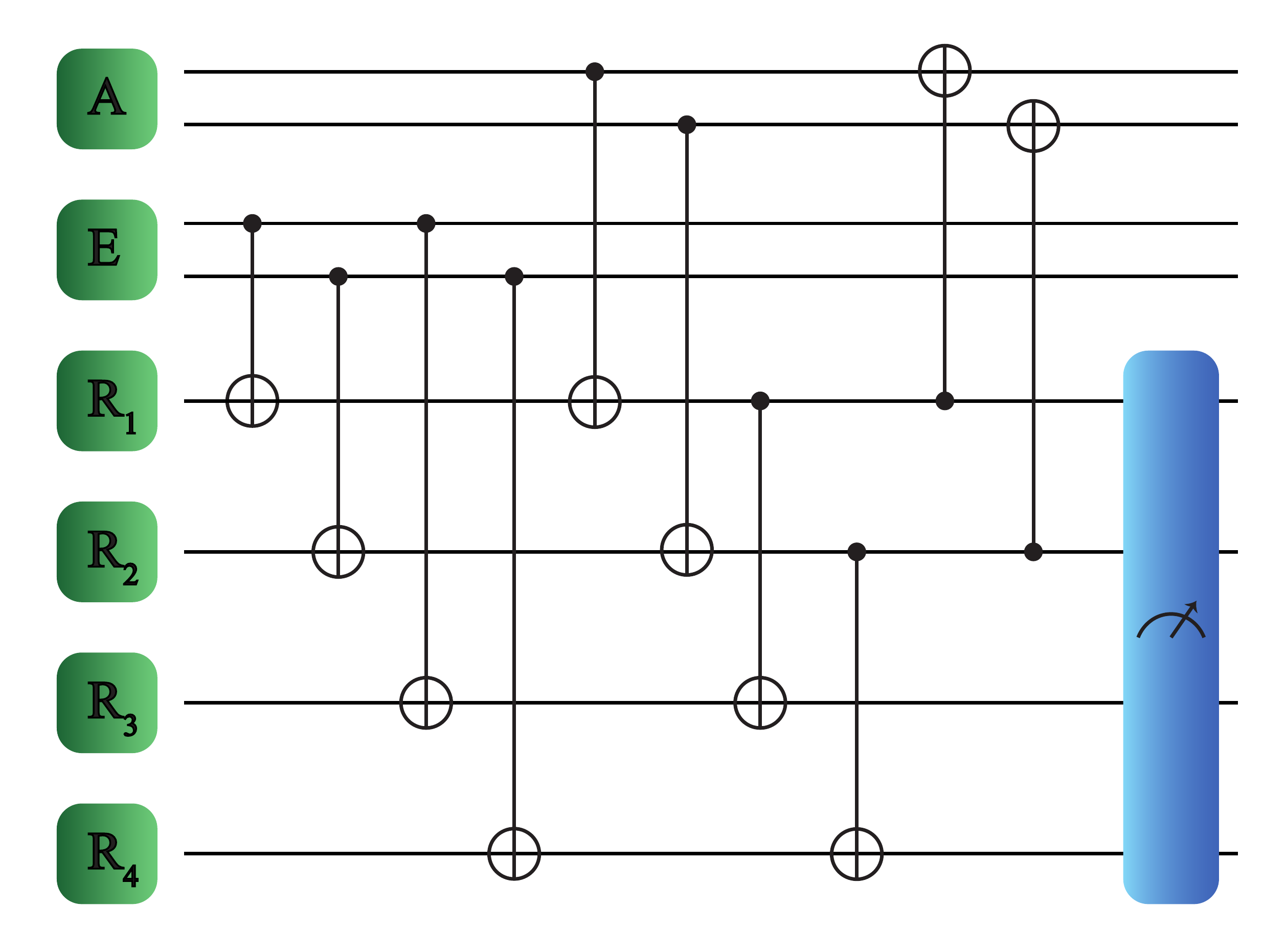}
\caption{\textbf{Schematic representation of quantum reinforcement learning protocol for multiqubit systems}. Agent, environment and registers are denoted as $\rm{A}$, $\rm{E}$ and $\rm{R}_1$, $\rm{R}_2$, $\rm{R}_3$ and $\rm{R}_4$, respectively. The measurement in the register subspace is denoted by the rightmost box.  }
\label{Fig2}
\end{figure}

In the previous section, we have showed that by considering more than just one register the rewarding criterion in the quantum reinforcement learning algorithm can be done at the end of our protocol. The same results can be obtained when we consider more complex configurations. Indeed, by assuming that agent and register are composed of two qubits each, and four qubits act as registers, we show that the rewarding criterion can also be applied at the end of the quantum protocol. Let us illustrate this fact with an analysis for multiqubit agent, environment, and register states,
\begin{eqnarray}
\label{eq6}
&&\ket{A}=\alpha_{A}^{00}\ket{00}_A + \alpha_{A}^{01}\ket{01}_A+ \alpha_{A}^{10}\ket{10}_A+ \alpha_{A}^{11}\ket{11}_A,\\
&&\ket{E} =\alpha_{E}^{00}\ket{00}_E + \alpha_{E}^{01}\ket{01}_E+ \alpha_{E}^{10}\ket{10}_E+ \alpha_{E}^{11}\ket{11}_E,\\
&&\ket{R} =\ket{0}_1\ket{0}_2\ket{0}_3\ket{0}_4,\\
&&\ket{\Psi}_0=\ket{A}\ket{E}\ket{R}.
\end{eqnarray}
Following the same procedure described previously, the protocol consists mainly in three types of interaction, as shown in \textbf{Fig. \ref{Fig2}}. Firstly, we update the registers conditionally to the environment states. More specifically, we consider an interaction between the environment qubits $E_1$ and $E_2$ with the registers $R_1$ and $R_2$, respectively. In this description, the environment acts as control and the registers act as targets in the CNOT gates,
\begin{eqnarray}
\label{eq9}\nonumber
\ket{\Psi}_1 & = & U^{\rm{CNOT}}_{({E_1},{R_1})}U^{\rm{CNOT}}_{({E_2},{R_2})},\ket{\Psi}_0,\\\nonumber
\ket{\Psi}_1 & = & \ket{A}\big(\alpha_{E}^{00}\ket{00}_E\ket{0}_1\ket{0}_2\ket{0}_3\ket{0}_4 + \alpha_{E}^{01}\ket{01}_E\ket{0}_1\ket{1}_2\ket{0}_3\ket{0}_4\\
&+& \alpha_{E}^{10}\ket{10}_E\ket{1}_1\ket{0}_2\ket{0}_3\ket{0}_4+ \alpha_{E}^{11}\ket{11}_E\ket{1}_1\ket{1}_2\ket{0}_3\ket{0}_4\big).
\end{eqnarray}
Thereafter, we update similarly the remaining registers, that is, we apply a CNOT gate between the environment qubits $E_1$ and $E_2$ and the register qubits $R_3$ and $R_4$, respectively, obtaining  
\begin{eqnarray}
\label{eq10}\nonumber
\ket{\Psi}_2 & = & U^{\rm{CNOT}}_{({E_1},{R_3})}U^{\rm{CNOT}}_{({E_2},{R_4})}\ket{\Psi}_1,\\\nonumber
\ket{\Psi}_2 & = & \ket{A}\big(\alpha_{E}^{00}\ket{00}_E\ket{0}_1\ket{0}_2\ket{0}_3\ket{0}_4 + \alpha_{E}^{01}\ket{01}_E\ket{0}_1\ket{1}_2\ket{0}_3\ket{1}_4\\
&+& \alpha_{E}^{10}\ket{10}_E\ket{1}_1\ket{0}_2\ket{1}_3\ket{0}_4+ \alpha_{E}^{11}\ket{11}_E\ket{1}_1\ket{1}_2\ket{1}_3\ket{1}_4\big).
\end{eqnarray}
Next step consists in updating a part of the register subspace conditionally to the agent state. Thus, the registers $R_1$ and $R_2$ will be updated via $A_1$ and $A_2$, respectively,
\begin{eqnarray}
\label{eq11}\nonumber
\ket{\Psi}_3 & = & U^{\rm{CNOT}}_{({A_1},{R_1})}U^{\rm{CNOT}}_{({A_2},{R_2})}\ket{\Psi}_2,\\\nonumber
\ket{\Psi}_3 & = &\alpha_{A}^{00}\alpha_{E}^{00}\ket{00}_A\ket{00}_E\ket{0}_1\ket{0}_2\ket{0}_3\ket{0}_4
+ \alpha_{A}^{00}\alpha_{E}^{01}\ket{00}_A\ket{01}_E\ket{0}_1\ket{1}_2\ket{0}_3\ket{1}_4\\\nonumber
&+& \alpha_{A}^{00}\alpha_{E}^{10}\ket{00}_A\ket{10}_E\ket{1}_1\ket{0}_2\ket{1}_3\ket{0}_4
+ \alpha_{A}^{00}\alpha_{E}^{11}\ket{00}_A\ket{11}_E\ket{1}_1\ket{1}_2\ket{1}_3\ket{1}_4\\\nonumber
&+&\alpha_{A}^{01}\alpha_{E}^{00}\ket{01}_A\ket{00}_E\ket{0}_1\ket{1}_2\ket{0}_3\ket{0}_4
+\alpha_{A}^{01}\alpha_{E}^{01}\ket{01}_A\ket{01}_E\ket{0}_1\ket{0}_2\ket{0}_3\ket{1}_4\\\nonumber
&+& \alpha_{A}^{01}\alpha_{E}^{10}\ket{01}_A\ket{10}_E\ket{1}_1\ket{1}_2\ket{1}_3\ket{0}_4
+ \alpha_{A}^{01}\alpha_{E}^{11}\ket{01}_A\ket{11}_E\ket{1}_1\ket{0}_2\ket{1}_3\ket{1}_4\\\nonumber
&+& \alpha_{A}^{10}\alpha_{E}^{00}\ket{10}_A\ket{00}_E\ket{1}_1\ket{0}_2\ket{0}_3\ket{0}_4
+\alpha_{A}^{10}\alpha_{E}^{01}\ket{10}_A\ket{01}_E\ket{1}_1\ket{1}_2\ket{0}_3\ket{1}_4\\\nonumber
&+& \alpha_{A}^{10}\alpha_{E}^{10}\ket{10}_A\ket{10}_E\ket{0}_1\ket{0}_2\ket{1}_3\ket{0}_4
+\alpha_{A}^{10}\alpha_{E}^{11}\ket{10}_A\ket{11}_E\ket{0}_1\ket{1}_2\ket{1}_3\ket{1}_4\\\nonumber
&+&\alpha_{A}^{11}\alpha_{E}^{00}\ket{11}_A\ket{00}_E\ket{1}_1\ket{1}_2\ket{0}_3\ket{0}_4
+ \alpha_{A}^{11}\alpha_{E}^{01}\ket{11}_A\ket{01}_E\ket{1}_1\ket{0}_2\ket{0}_3\ket{1}_4\\
&+& \alpha_{A}^{11}\alpha_{E}^{10}\ket{11}_A\ket{10}_E\ket{0}_1\ket{1}_2\ket{1}_3\ket{0}_4
+\alpha_{A}^{11}\alpha_{E}^{11}\ket{11}_A\ket{11}_E\ket{0}_1\ket{0}_2\ket{1}_3\ket{1}_4.
\end{eqnarray}
Afterwards, to obtain orthogonal outcomes in the register subspace we perform a pair of CNOT gates in this subspace. The interaction will be between the registers that interact with a common environment, namely, register $R_1$ interacts with $R_3$ because both have interacted with $E_1$. Similarly for $R_2$ and $R_4$, which have interacted with $E_2$. In this case, $R_1$($R_2$) is the control and $R_3$($R_4$) is the target.
\begin{eqnarray}
\label{eq12}\nonumber
\ket{\Psi}_4 & = & U^{\rm{CNOT}}_{({R_1},{R_3})}U^{\rm{CNOT}}_{({R_2},{R_4})}\ket{\Psi}_3,\\\nonumber
\ket{\Psi}_4 & = &\alpha_{A}^{00}\alpha_{E}^{00}\ket{00}_A\ket{00}_E\ket{0}_1\ket{0}_2\ket{0}_3\ket{0}_4
+ \alpha_{A}^{00}\alpha_{E}^{01}\ket{00}_A\ket{01}_E\ket{0}_1\ket{1}_2\ket{0}_3\ket{0}_4\\\nonumber
&+& \alpha_{A}^{00}\alpha_{E}^{10}\ket{00}_A\ket{10}_E\ket{1}_1\ket{0}_2\ket{0}_3\ket{0}_4
+ \alpha_{A}^{00}\alpha_{E}^{11}\ket{00}_A\ket{11}_E\ket{1}_1\ket{1}_2\ket{0}_3\ket{0}_4 \\\nonumber 
&+&\alpha_{A}^{01}\alpha_{E}^{00}\ket{01}_A\ket{00}_E\ket{0}_1\ket{1}_2\ket{0}_3\ket{1}_4
+\alpha_{A}^{01}\alpha_{E}^{01}\ket{01}_A\ket{01}_E\ket{0}_1\ket{0}_2\ket{0}_3\ket{1}_4\\\nonumber
&+& \alpha_{A}^{01}\alpha_{E}^{10}\ket{01}_A\ket{10}_E\ket{1}_1\ket{1}_2\ket{0}_3\ket{1}_4
+ \alpha_{A}^{01}\alpha_{E}^{11}\ket{01}_A\ket{11}_E\ket{1}_1\ket{0}_2\ket{0}_3\ket{1}_4\\\nonumber
&+& \alpha_{A}^{10}\alpha_{E}^{00}\ket{10}_A\ket{00}_E\ket{1}_1\ket{0}_2\ket{1}_3\ket{0}_4
+\alpha_{A}^{10}\alpha_{E}^{01}\ket{10}_A\ket{01}_E\ket{1}_1\ket{1}_2\ket{1}_3\ket{0}_4\\\nonumber
&+& \alpha_{A}^{10}\alpha_{E}^{10}\ket{10}_A\ket{10}_E\ket{0}_1\ket{0}_2\ket{1}_3\ket{0}_4
+\alpha_{A}^{10}\alpha_{E}^{11}\ket{10}_A\ket{11}_E\ket{0}_1\ket{1}_2\ket{1}_3\ket{0}_4\\\nonumber
&+&\alpha_{A}^{11}\alpha_{E}^{00}\ket{11}_A\ket{00}_E\ket{1}_1\ket{1}_2\ket{1}_3\ket{1}_4
+ \alpha_{A}^{11}\alpha_{E}^{01}\ket{11}_A\ket{01}_E\ket{1}_1\ket{0}_2\ket{1}_3\ket{1}_4\\
&+& \alpha_{A}^{11}\alpha_{E}^{10}\ket{11}_A\ket{10}_E\ket{0}_1\ket{1}_2\ket{1}_3\ket{1}_4
+\alpha_{A}^{11}\alpha_{E}^{11}\ket{11}_A\ket{11}_E\ket{0}_1\ket{0}_2\ket{1}_3\ket{1}_4.
\end{eqnarray}
Finally, we update the agent considering the states of the register in order that the rewarding criterion is satisfied. This is done by applying two CNOT gates in the agent-register subspace, where $A_1$ is controlled by $R_1$ and $A_2$ is controlled by $R_2$, 
\begin{eqnarray}\nonumber
\ket{\Psi}_5 & = & U^{\rm{CNOT}}_{({R_1},{A_1})}U^{\rm{CNOT}}_{({R_2},{A_2})}\ket{\Psi}_4,\\\nonumber
\label{eq13}
\ket{\Psi}_5 & = &\alpha_{A}^{00}\alpha_{E}^{00}\ket{00}_A\ket{00}_E\ket{0}_1\ket{0}_2\ket{0}_3\ket{0}_4
+ \alpha_{A}^{00}\alpha_{E}^{01}\ket{01}_A\ket{01}_E\ket{0}_1\ket{1}_2\ket{0}_3\ket{0}_4\\\nonumber
&+&\alpha_{A}^{00}\alpha_{E}^{10}\ket{10}_A\ket{10}_E\ket{1}_1\ket{0}_2\ket{0}_3\ket{0}_4
+ \alpha_{A}^{00}\alpha_{E}^{11}\ket{11}_A\ket{11}_E\ket{1}_1\ket{1}_2\ket{0}_3\ket{0}_4\\\nonumber
&+&\alpha_{A}^{01}\alpha_{E}^{00}\ket{00}_A\ket{00}_E\ket{0}_1\ket{1}_2\ket{0}_3\ket{1}_4
+\alpha_{A}^{01}\alpha_{E}^{01}\ket{01}_A\ket{01}_E\ket{0}_1\ket{0}_2\ket{0}_3\ket{1}_4\\\nonumber
&+&\alpha_{A}^{01}\alpha_{E}^{10}\ket{10}_A\ket{10}_E\ket{1}_1\ket{1}_2\ket{0}_3\ket{1}_4
+ \alpha_{A}^{01}\alpha_{E}^{11}\ket{11}_A\ket{11}_E\ket{1}_1\ket{0}_2\ket{0}_3\ket{1}_4\\\nonumber
&+&\alpha_{A}^{10}\alpha_{E}^{00}\ket{00}_A\ket{00}_E\ket{1}_1\ket{0}_2\ket{1}_3\ket{0}_4
+\alpha_{A}^{10}\alpha_{E}^{01}\ket{01}_A\ket{01}_E\ket{1}_1\ket{1}_2\ket{1}_3\ket{0}_4\\\nonumber
&+&\alpha_{A}^{10}\alpha_{E}^{10}\ket{10}_A\ket{10}_E\ket{0}_1\ket{0}_2\ket{1}_3\ket{0}_4
+\alpha_{A}^{10}\alpha_{E}^{11}\ket{11}_A\ket{11}_E\ket{0}_1\ket{1}_2\ket{1}_3\ket{0}_4\\\nonumber
&+&\alpha_{A}^{11}\alpha_{E}^{00}\ket{00}_A\ket{00}_E\ket{1}_1\ket{1}_2\ket{1}_3\ket{1}_4
+ \alpha_{A}^{11}\alpha_{E}^{01}\ket{01}_A\ket{01}_E\ket{1}_1\ket{0}_2\ket{1}_3\ket{1}_4\\
&+&\alpha_{A}^{11}\alpha_{E}^{10}\ket{10}_A\ket{10}_E\ket{0}_1\ket{1}_2\ket{1}_3\ket{1}_4
+\alpha_{A}^{11}\alpha_{E}^{11}\ket{11}_A\ket{11}_E\ket{0}_1\ket{0}_2\ket{1}_3\ket{1}_4.
\end{eqnarray}
From the latter Eq.~(\ref{eq13}), it is straightforward to see that independently of the measurement outcomes the learning fidelity is maximal. Moreover, as in the previous case, one iteration of the quantum reinforcement protocol is needed to obtain maximal learning fidelity, $\mathcal{F}_{AE} = |\braket{\psi_A}{\phi_E}|$.\par

\section*{Quantum reinforcement learning for qudit systems}  
So far, we have studied quantum reinforcement learning processes only for two-level systems or in pairs of them. However, there are several quantum systems which cannot be described in terms of a two-level system. For instance, quantum harmonic oscillators, electronic energy levels in an ion, and superconducting artificial atoms such as transmons~\cite{PhysRevA.76.042319}, where for some regimes of Josephson energy they must be considered as a three-level system. In this context, it is interesting to extend the quantum reinforcement learning protocol developed here for cases where multilevel systems compound the agent, environment, and register.

To perform the previous task, we first need to define a set of logic operations that we will perform on our system. In the qubit case, the main logical operation applied is the CNOT gate, which considers a conditional interaction between two qubits, where one acts as a control while the other acts as a target. The control qubit remains unchanged whereas the target qubit output is modified by the addition modulo 2. Then, it is wise to assume that the set of logic operations between multilevel systems could be defined in terms of an addition modulo $\mathcal{D}$, where $\mathcal{D}$ stands for the dimension of one subsystem (agent, environment or register subspaces), according to
\begin{eqnarray}
\label{eq14}
U\ket{i}_1\ket{j}_{2} = \ket{i}_{1}\ket{i \oplus j}_{2}.
\end{eqnarray}
Here, $i \oplus j$ stands for the addition modulo $\mathcal{D}$. This gate is usually known as \textit{XOR} gate \cite{alber2000generalized}. For two-dimensional systems, this gate corresponds to the CNOT gate. Nevertheless, for higher dimensional systems this definition presents several disadvantages. For instance, the \textit{XOR} gate defined as in Eq.~(\ref{eq14}) is unitary but not Hermitian for $\mathcal{D}>2$. Moreover, this logical operation is no longer its own inverse. To avoid these problems, in the literature~\cite{alber2000generalized} the generalized XOR gate (GXOR) has been defined as
\begin{eqnarray}
\label{eq15}
\text{GXOR}_{1,2}\ket{i}_1\ket{j}_{2} = \ket{i}_{1}\ket{i \ominus j}_2,
\end{eqnarray}
where the operation $\ominus$ denotes the difference $i-j$ \textit{modulo} $\mathcal{D}$. The GXOR gate of Eq.~(\ref{eq15}) does not present the disadvantages pointed out in the definition of Eq. (\ref{eq14}). That is, the GXOR gate is Hermitian, unitary and $i\ominus j=0$ only when $i=j$.\par
Considering our proposed protocol for single-qubit cases, we show that when we take into account multilevel systems, the number of interactions to obtain maximal learning fidelity is fixed and depends only on the number of agent subsystems in the protocol. Let us illustrate this with an example of multilevel agent-environment-register state,
\begin{eqnarray}
\label{eq28}
\ket{\Psi_0}&=&\sum_{n=0}^{N-1}\sum_{m=0}^{N-1}\alpha_{A}^{n}\alpha_{E}^{m}\ket{n}_A\ket{m}_E\ket{0}_1\ket{0}_2.
\end{eqnarray} 
The first step in our protocol is identical to the equivalent one in the single-qubit case. We update the register conditionally on the environment state, that is, we transfer information of the environment and encode it in the register system. This is done by applying a pair of GXOR gates acting in the environment-register subsystem. In this case, the environment interacts with both registers $R_1$ and $R_2$.  The environment acts as control and both registers are targets,
\begin{eqnarray}
\label{eq29}\nonumber
\ket{\Psi_1}&= &   U^{\rm{GXOR}}_{({E},{R_1})}\ket{\Psi_0},\\
\ket{\Psi_1} & = & \sum_{n=0}^{N-1}\sum_{m=0}^{N-1}\alpha_{A}^{n}\alpha_{E}^{m}\ket{n}_A\ket{m}_E\ket{m}_1\ket{0}_2.
\end{eqnarray}

\begin{eqnarray}
\label{eq29}\nonumber
\ket{\Psi_2}&= &   U^{\rm{GXOR}}_{({E},{R_2})}\ket{\Psi_1},\\
\ket{\Psi_2} & = & \sum_{n=0}^{N-1}\sum_{m=0}^{N-1}\alpha_{A}^{n}\alpha_{E}^{m}\ket{n}_A\ket{m}_E\ket{m}_1\ket{m}_2.
\end{eqnarray}

\begin{figure}[!h]
\centering
\includegraphics[width=1\linewidth,height=6cm,keepaspectratio]{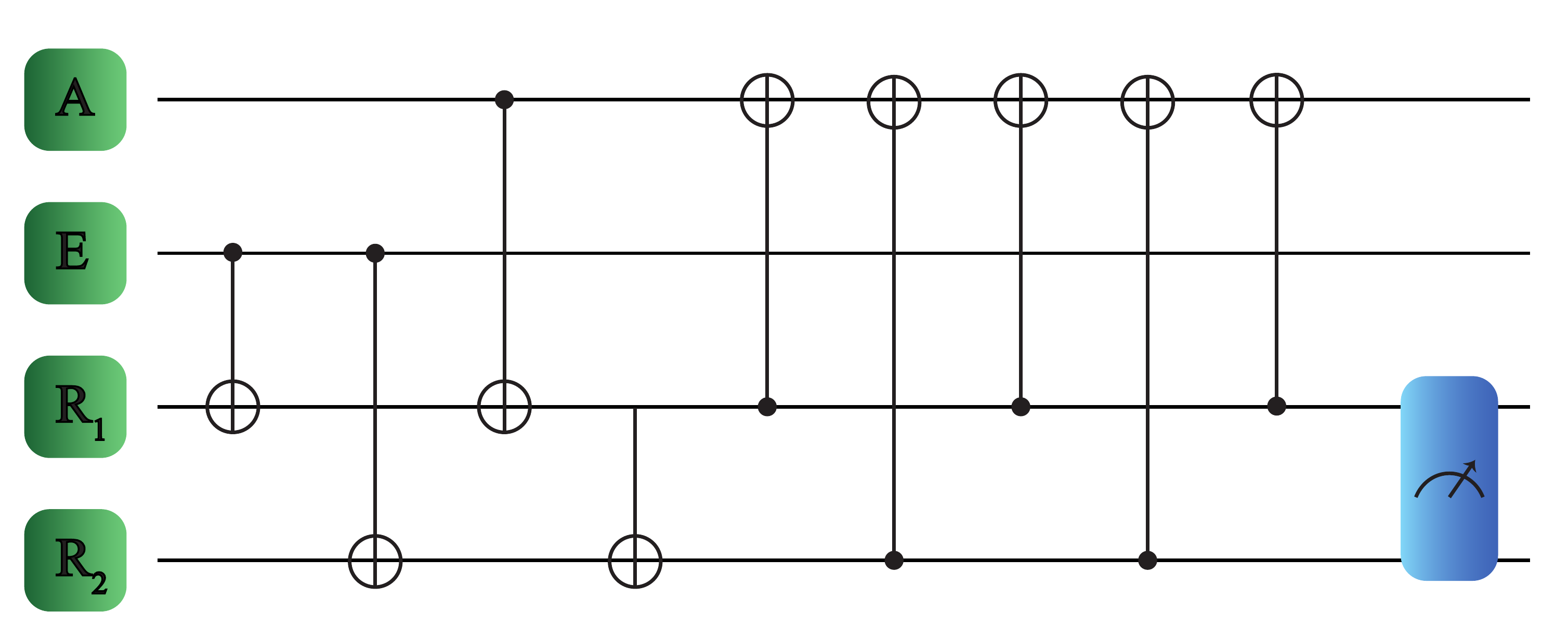}
\caption{\textbf{Quantum reinforcement learning protocol for qudits}. The systems involved are denoted as agent A, environment E and registers $R_1,R_2$. In this case, the logical quantum gates which are applied in the learning protocol correspond to GXOR gates. The measurement process in the register subspace is denoted with the rightmost box.}
\label{Fig3}
\end{figure}

Once the information has been transferred to the register, we update the register $R_1$ based on the agent state. That is, we perform a GXOR gate in the subspace composed of agent and register. Here, the agent act as a control and the register $R_1$ is the target,
\begin{eqnarray}
\label{eq30}\nonumber
\ket{\Psi_3}&&=  U^{\rm{GXOR}}_{(A,{R_1})}\ket{\Psi_2},\\
\ket{\Psi_3}&&=\sum_{n=0}^{N-1}\sum_{m=0}^{N-1}\alpha_{A}^{n}\alpha_{E}^{m}\ket{n}_A\ket{m}_E\ket{n\ominus m}_1\ket{m}_2.
\end{eqnarray}
Orthogonal outcome measurements in the register subspace are provided by interactions between the registers in this subspace. Thus, we apply a GXOR gate in the register subspace, where $R_1$ is the control and $R_2$ is the target,
\begin{eqnarray}
\label{eq31}\nonumber
\ket{\Psi_4}&&=  U^{\rm{GXOR}}_{(R_1,{R_2})}\ket{\Psi_3},\\
\ket{\Psi_4}&&=\sum_{n=0}^{N-1}\sum_{m=0}^{N-1}\alpha_{A}^{n}\alpha_{E}^{m}\ket{n}_A\ket{m}_E\ket{n\ominus m}_1\ket{(n\ominus m)\ominus m}_2.
\end{eqnarray}
Subsequently, the agent state is updated conditionally to the information encoded in the state of the register $R_1$. The GXOR gate is applied in the register-agent subspace. In this case, $R_1$ is the control and the agent is the target,
\begin{eqnarray}
\label{eq32}\nonumber
\ket{\Psi_5}&&=  U^{\rm{GXOR}}_{({R_1},A)}\ket{\Psi_4},\\
\ket{\Psi_5}&&=\sum_{n=0}^{N-1}\sum_{m=0}^{N-1}\alpha_{A}^{n}\alpha_{E}^{m}\ket{0\ominus m}_A\ket{m}_E\ket{n\ominus m}_1\ket{n\ominus 2m}_2.
\end{eqnarray}
For the case where the multi-level system contains $\mathcal{D}=2$, we recover the result discussed previously because of $0 \ominus m=m$ for that dimension. On the other hand, we are interested in systems with more energy levels, such that we need to adapt the protocol to obtain maximal learning fidelity for a fixed number of steps. In this case, we will update the agent subsystem by an iterative interaction with registers $R_1$ and $R_2$ as shown in \textbf{Fig. \ref{Fig3}}. Here, the agent always acts as target, while the registers are the controls. Therefore, we apply a GXOR gate between the register $R_2$ and the agent,
\begin{eqnarray}
\label{eq33}\nonumber
\ket{\Psi_6} & = &  U^{\rm{GXOR}}_{({R_2},A)}\ket{\Psi_5},\\
\ket{\Psi_6} & = & \sum_{n=0}^{N-1}\sum_{m=0}^{N-1}\alpha_{A}^{n}\alpha_{E}^{m}\ket{n\ominus m}_A\ket{m}_E\ket{n\ominus m}_1\ket{n\ominus 2m}_2.
\end{eqnarray}
Now, by applying a GXOR gate between the register $R_1$ and the agent we obtain,
\begin{eqnarray}
\label{eq34}\nonumber
\ket{\Psi_7} & = &  U^{\rm{GXOR}}_{({R_1},A)}\ket{\Psi_6},\\
\ket{\Psi_7} & = & \sum_{n=0}^{N-1}\sum_{m=0}^{N-1}\alpha_{A}^{n}\alpha_{E}^{m}\ket{0}_A\ket{m}_E\ket{n\ominus m}_1\ket{n\ominus 2m}_2.
\end{eqnarray}
We perform subsequently a GXOR gate in the subspace composed of $R_2$ and agent $A$,
\begin{eqnarray}
\label{eq35}\nonumber
\ket{\Psi_8} & = &  U^{\rm{GXOR}}_{({R_2},A)}\ket{\Psi_7},\\
\ket{\Psi_8} & = & \sum_{n=0}^{N-1}\sum_{m=0}^{N-1}\alpha_{A}^{n}\alpha_{E}^{m}\ket{n\ominus 2m}_A\ket{m}_E\ket{n\ominus m}_1\ket{n\ominus 2m}_2.
\end{eqnarray}
Finally, applying a GXOR gate on the register-agent subspace we obtain the desired result. By considering a fixed number of interactions between the set of agent, environment and register, the learning fidelity becomes maximal independently of the outcome measurement on the register subspace, which can again be carried out at the end of the protocol,
\begin{eqnarray}
\label{eq36}\nonumber
\ket{\Psi_9} & = &  U^{\rm{GXOR}}_{({R_1},A)}\ket{\Psi_8},\\
\ket{\Psi_9} &  = & \sum_{n=0}^{N-1}\sum_{m=0}^{N-1}\alpha_{A}^{n}\alpha_{E}^{m}\ket{m}_A\ket{m}_E\ket{n\ominus m}_1\ket{n\ominus 2m}_2.
\end{eqnarray}
Thus, in a machine learning protocol where the learning units are composed by multilevel systems (see Fig \ref{Fig3}), the number of logical operations required to obtain maximal learning fidelity does not depend on the system dimension. 
\subsection*{Example}
Here, we exemplify how our reinforcement learning protocol works in qudit systems. We consider, without loss of generality, the case for dimension $\mathcal{D}=4$. In this case, the agent-environment-register state has the following form,
\begin{eqnarray}
\label{eq21}
&&\ket{A} = \alpha_{A}^{0}\ket{0}_{A} + \alpha_{A}^{1}\ket{1}_{A} + \alpha_{A}^{2}\ket{2}_{A} + \alpha_{A}^{3}\ket{3}_{A}\\,
&&\ket{E} = \alpha_{E}^{0}\ket{0}_{E} + \alpha_{E}^{1}\ket{1}_{E} + \alpha_{E}^{2}\ket{2}_{E} + \alpha_{E}^{3}\ket{3}_{E}\\
&&\ket{R} = \ket{0}_1\ket{0}_2\\
&&\ket{\Psi}_0=\ket{A}\ket{E}\ket{R}.
\end{eqnarray}
As mentioned previously, the considered quantum gate is a GXOR gate with subtraction modulo 4. The first step is to update the register according to the environment information,
\begin{eqnarray}
\label{eq22a}\nonumber
\ket{\Psi}_1  & = & U^{\rm{GXOR}}_{({E},{R_1})}\ket{\Psi}_0,\\\nonumber
\ket{\Psi}_1  & = & ( \alpha_{A}^{0}\ket{0}_{A} + \alpha_{A}^{1}\ket{1}_{A} + \alpha_{A}^{2}\ket{2}_{A} + \alpha_{A}^{3}\ket{3}_{A})\\\nonumber
&&(
\alpha_{E}^{0}\ket{0}_E\ket{0}_1\ket{0}_2
+\alpha_{E}^{1}\ket{1}_E\ket{1}_1\ket{0}_2
+\alpha_{E}^{2}\ket{2}_E\ket{2}_1\ket{0}_2
+\alpha_{E}^{3}\ket{3}_E\ket{3}_1\ket{0}_2 ),\\
\end{eqnarray} 
\begin{eqnarray}
\label{eq22b}\nonumber
\ket{\Psi}_2  & = & U^{\rm{GXOR}}_{({E},{R_2})}\ket{\Psi}_1,\\\nonumber
\ket{\Psi}_2  & = & ( \alpha_{A}^{0}\ket{0}_{A} + \alpha_{A}^{1}\ket{1}_{A} + \alpha_{A}^{2}\ket{2}_{A} + \alpha_{A}^{3}\ket{3}_{A})\\\nonumber
&&(
\alpha_{E}^{0}\ket{0}_E\ket{0}_1\ket{0}_2
+\alpha_{E}^{1}\ket{1}_E\ket{1}_1\ket{1}_2
+\alpha_{E}^{2}\ket{2}_E\ket{2}_1\ket{2}_2
+\alpha_{E}^{3}\ket{3}_E\ket{3}_1\ket{3}_2 ).\\
\end{eqnarray} 
Subsequently, the register is updated conditional to the agent state,
\begin{eqnarray}
\label{eq23}\nonumber
\ket{\Psi}_3  & = & U^{\rm{GXOR}}_{({A},{R_1})}\ket{\Psi}_2,\\\nonumber
\ket{\Psi}_3  & = &\alpha_{A}^{0}\alpha_{E}^{0}\ket{0}_A\ket{0}_E\ket{0}_1\ket{0}_2
+ \alpha_{A}^{0}\alpha_{E}^{1}\ket{0}_A\ket{1}_E\ket{3}_1\ket{1}_2
+ \alpha_{A}^{0}\alpha_{E}^{2}\ket{0}_A\ket{2}_E\ket{2}_1\ket{2}_2\\\nonumber
&+& \alpha_{A}^{0}\alpha_{E}^{3}\ket{0}_A\ket{3}_E\ket{1}_1\ket{3}_2+ \alpha_{A}^{1}\alpha_{E}^{0}\ket{1}_A\ket{0}_E\ket{1}_1\ket{0}_2
+ \alpha_{A}^{1}\alpha_{E}^{1}\ket{1}_A\ket{1}_E\ket{0}_1\ket{1}_2\\\nonumber
&+&\alpha_{A}^{1}\alpha_{E}^{2}\ket{1}_A\ket{2}_E\ket{3}_1\ket{2}_2
+\alpha_{A}^{1}\alpha_{E}^{3}\ket{1}_A\ket{3}_E\ket{2}_1\ket{3}_2
+ \alpha_{A}^{2}\alpha_{E}^{0}\ket{2}_A\ket{0}_E\ket{2}_1\ket{0}_2\\\nonumber
&+& \alpha_{A}^{2}\alpha_{E}^{1}\ket{2}_A\ket{1}_E\ket{1}_1\ket{1}_2
+\alpha_{A}^{2}\alpha_{E}^{2}\ket{2}_A\ket{2}_E\ket{0}_1\ket{2}_2
+\alpha_{A}^{2}\alpha_{E}^{3}\ket{2}_A\ket{3}_E\ket{3}_1\ket{3}_2\\\nonumber
&+&\alpha_{A}^{3}\alpha_{E}^{0}\ket{3}_A\ket{0}_E\ket{3}_1\ket{0}_2
+ \alpha_{A}^{3}\alpha_{E}^{1}\ket{3}_A\ket{1}_E\ket{2}_1\ket{1}_2
+\alpha_{A}^{3}\alpha_{E}^{2}\ket{3}_A\ket{2}_E\ket{1}_1\ket{2}_2\\
&+&\alpha_{A}^{3}\alpha_{E}^{3}\ket{3}_A\ket{3}_E\ket{0}_1\ket{3}_2.
\end{eqnarray}
Then, to obtain orthogonal outcome measurements in the register basis, we perform an interaction in the register subspace,
\begin{eqnarray}
\label{eq24}\nonumber
\ket{\Psi}_4  & = &  U^{\rm{GXOR}}_{({R_1},{R_2})}\ket{\Psi}_3,\\\nonumber
\ket{\Psi}_4  & = &\alpha_{A}^{0}\alpha_{E}^{0}\ket{0}_A\ket{0}_E\ket{0}_1\ket{0}_2
+ \alpha_{A}^{0}\alpha_{E}^{1}\ket{0}_A\ket{1}_E\ket{3}_1\ket{2}_2
+ \alpha_{A}^{0}\alpha_{E}^{2}\ket{0}_A\ket{2}_E\ket{2}_1\ket{0}_2\\\nonumber
&+& \alpha_{A}^{0}\alpha_{E}^{3}\ket{0}_A\ket{3}_E\ket{1}_1\ket{2}_2+ \alpha_{A}^{1}\alpha_{E}^{0}\ket{1}_A\ket{0}_E\ket{1}_1\ket{1}_2
+ \alpha_{A}^{1}\alpha_{E}^{1}\ket{1}_A\ket{1}_E\ket{0}_1\ket{3}_2\\\nonumber
&+&\alpha_{A}^{1}\alpha_{E}^{2}\ket{1}_A\ket{2}_E\ket{3}_1\ket{1}_2
+\alpha_{A}^{1}\alpha_{E}^{3}\ket{1}_A\ket{3}_E\ket{2}_1\ket{3}_2
+\alpha_{A}^{2}\alpha_{E}^{0}\ket{2}_A\ket{0}_E\ket{2}_1\ket{2}_2\\\nonumber
&+& \alpha_{A}^{2}\alpha_{E}^{1}\ket{2}_A\ket{1}_E\ket{1}_1\ket{0}_2
+\alpha_{A}^{2}\alpha_{E}^{2}\ket{2}_A\ket{2}_E\ket{0}_1\ket{2}_2
+\alpha_{A}^{2}\alpha_{E}^{3}\ket{2}_A\ket{3}_E\ket{3}_1\ket{0}_2\\\nonumber
&+&\alpha_{A}^{3}\alpha_{E}^{0}\ket{3}_A\ket{0}_E\ket{3}_1\ket{3}_2
+ \alpha_{A}^{3}\alpha_{E}^{1}\ket{3}_A\ket{1}_E\ket{2}_1\ket{1}_2
+\alpha_{A}^{3}\alpha_{E}^{2}\ket{3}_A\ket{2}_E\ket{1}_1\ket{3}_2\\
&+&\alpha_{A}^{3}\alpha_{E}^{3}\ket{3}_A\ket{3}_E\ket{0}_1\ket{1}_2.
\end{eqnarray}
Now, we need to apply iterative interactions in the register-agent subspace to update the agent in each step until we get maximal learning fidelity with respect to the environment. We start by performing a GXOR gate between the register $R_1$ and the agent,
\begin{eqnarray}
\label{eq25}\nonumber
\ket{\Psi}_5  & = & U^{\rm{GXOR}}_{({R_1},{A})}\ket{\Psi}_4,\\\nonumber
\ket{\Psi}_5  & = &\alpha_{A}^{0}\alpha_{E}^{0}\ket{0}_A\ket{0}_E\ket{0}_1\ket{0}_2
+ \alpha_{A}^{0}\alpha_{E}^{1}\ket{3}_A\ket{1}_E\ket{3}_1\ket{2}_2
+ \alpha_{A}^{0}\alpha_{E}^{2}\ket{2}_A\ket{2}_E\ket{2}_1\ket{0}_2\\\nonumber
&+& \alpha_{A}^{0}\alpha_{E}^{3}\ket{1}_A\ket{3}_E\ket{1}_1\ket{2}_2
+ \alpha_{A}^{1}\alpha_{E}^{0}\ket{0}_A\ket{0}_E\ket{1}_1\ket{1}_2
+ \alpha_{A}^{1}\alpha_{E}^{1}\ket{3}_A\ket{1}_E\ket{0}_1\ket{3}_2\\\nonumber
&+&\alpha_{A}^{1}\alpha_{E}^{2}\ket{2}_A\ket{2}_E\ket{3}_1\ket{1}_2
+\alpha_{A}^{1}\alpha_{E}^{3}\ket{1}_A\ket{3}_E\ket{2}_1\ket{3}_2
+\alpha_{A}^{2}\alpha_{E}^{0}\ket{0}_A\ket{0}_E\ket{2}_1\ket{2}_2\\\nonumber
&+& \alpha_{A}^{2}\alpha_{E}^{1}\ket{3}_A\ket{1}_E\ket{1}_1\ket{0}_2
+\alpha_{A}^{2}\alpha_{E}^{2}\ket{2}_A\ket{2}_E\ket{0}_1\ket{2}_2
+\alpha_{A}^{2}\alpha_{E}^{3}\ket{1}_A\ket{3}_E\ket{3}_1\ket{0}_2\\\nonumber
&+&\alpha_{A}^{3}\alpha_{E}^{0}\ket{0}_A\ket{0}_E\ket{3}_1\ket{3}_2
+ \alpha_{A}^{3}\alpha_{E}^{1}\ket{3}_A\ket{1}_E\ket{2}_1\ket{1}_2
+\alpha_{A}^{3}\alpha_{E}^{2}\ket{2}_A\ket{2}_E\ket{1}_1\ket{3}_2\\
&+&\alpha_{A}^{3}\alpha_{E}^{3}\ket{1}_A\ket{3}_E\ket{0}_1\ket{1}_2.
\end{eqnarray}	
Hereafter, we apply the GXOR gate in the $R_2$-agent subspace,	
\begin{eqnarray}
\label{eq26}\nonumber
\ket{\Psi}_6  & = & U^{\rm{GXOR}}_{({R_2},{A})}\ket{\Psi}_5,\\\nonumber
\ket{\Psi}_6  & = &\alpha_{A}^{0}\alpha_{E}^{0}\ket{0}_A\ket{0}_E\ket{0}_1\ket{0}_2
+ \alpha_{A}^{0}\alpha_{E}^{1}\ket{3}_A\ket{1}_E\ket{3}_1\ket{2}_2
+ \alpha_{A}^{0}\alpha_{E}^{2}\ket{2}_A\ket{2}_E\ket{2}_1\ket{0}_2\\\nonumber
&+& \alpha_{A}^{0}\alpha_{E}^{3}\ket{1}_A\ket{3}_E\ket{1}_1\ket{2}_2
+ \alpha_{A}^{1}\alpha_{E}^{0}\ket{1}_A\ket{0}_E\ket{1}_1\ket{1}_2
+ \alpha_{A}^{1}\alpha_{E}^{1}\ket{0}_A\ket{1}_E\ket{0}_1\ket{3}_2\\\nonumber
&+&\alpha_{A}^{1}\alpha_{E}^{2}\ket{3}_A\ket{2}_E\ket{3}_1\ket{1}_2
+\alpha_{A}^{1}\alpha_{E}^{3}\ket{2}_A\ket{3}_E\ket{2}_1\ket{3}_2
+ \alpha_{A}^{2}\alpha_{E}^{0}\ket{2}_A\ket{0}_E\ket{2}_1\ket{2}_2\\\nonumber
&+& \alpha_{A}^{2}\alpha_{E}^{1}\ket{1}_A\ket{1}_E\ket{1}_1\ket{0}_2
+\alpha_{A}^{2}\alpha_{E}^{2}\ket{0}_A\ket{2}_E\ket{0}_1\ket{2}_2
+\alpha_{A}^{2}\alpha_{E}^{3}\ket{3}_A\ket{3}_E\ket{3}_1\ket{0}_2\\\nonumber
&+&\alpha_{A}^{3}\alpha_{E}^{0}\ket{3}_A\ket{0}_E\ket{3}_1\ket{3}_2
+ \alpha_{A}^{3}\alpha_{E}^{1}\ket{2}_A\ket{1}_E\ket{2}_1\ket{1}_2
+\alpha_{A}^{3}\alpha_{E}^{2}\ket{1}_A\ket{2}_E\ket{1}_1\ket{3}_2\\
&+&\alpha_{A}^{3}\alpha_{E}^{3}\ket{0}_A\ket{3}_E\ket{0}_1\ket{1}_2.
\end{eqnarray}
Afterwards, we perform a GXOR gate between $R_1$ and A,
\begin{eqnarray}
\label{eq27}\nonumber
\ket{\Psi}_7  & = & U^{\rm{GXOR}}_{({R_1},{A})}\ket{\Psi}_6,\\\nonumber
\ket{\Psi}_7  & = &\alpha_{A}^{0}\alpha_{E}^{0}\ket{0}_A\ket{0}_E\ket{0}_1\ket{0}_2
+ \alpha_{A}^{0}\alpha_{E}^{1}\ket{0}_A\ket{1}_E\ket{3}_1\ket{2}_2
+ \alpha_{A}^{0}\alpha_{E}^{2}\ket{0}_A\ket{2}_E\ket{2}_1\ket{0}_2\\\nonumber
&+& \alpha_{A}^{0}\alpha_{E}^{3}\ket{0}_A\ket{3}_E\ket{1}_1\ket{2}_2
+ \alpha_{A}^{1}\alpha_{E}^{0}\ket{0}_A\ket{0}_E\ket{1}_1\ket{1}_2
+ \alpha_{A}^{1}\alpha_{E}^{1}\ket{0}_A\ket{1}_E\ket{0}_1\ket{3}_2\\\nonumber
&+&\alpha_{A}^{1}\alpha_{E}^{2}\ket{0}_A\ket{2}_E\ket{3}_1\ket{1}_2
+\alpha_{A}^{1}\alpha_{E}^{3}\ket{0}_A\ket{3}_E\ket{2}_1\ket{3}_2
+ \alpha_{A}^{2}\alpha_{E}^{0}\ket{0}_A\ket{0}_E\ket{2}_1\ket{2}_2\\\nonumber
&+& \alpha_{A}^{2}\alpha_{E}^{1}\ket{0}_A\ket{1}_E\ket{1}_1\ket{0}_2
+\alpha_{A}^{2}\alpha_{E}^{2}\ket{0}_A\ket{2}_E\ket{0}_1\ket{2}_2
+\alpha_{A}^{2}\alpha_{E}^{3}\ket{0}_A\ket{3}_E\ket{3}_1\ket{0}_2\\\nonumber
&+&\alpha_{A}^{3}\alpha_{E}^{0}\ket{0}_A\ket{0}_E\ket{3}_1\ket{3}_2
+ \alpha_{A}^{3}\alpha_{E}^{1}\ket{0}_A\ket{1}_E\ket{2}_1\ket{1}_2
+\alpha_{A}^{3}\alpha_{E}^{2}\ket{0}_A\ket{2}_E\ket{1}_1\ket{3}_2\\
&+&\alpha_{A}^{3}\alpha_{E}^{3}\ket{0}_A\ket{3}_E\ket{0}_1\ket{1}_2.
\end{eqnarray}
Subsequently, an interaction in the $R_2$-agent subspace is performed,
\begin{eqnarray}
\label{eq28b}\nonumber
\ket{\Psi}_8  & = & U^{\rm{GXOR}}_{({R_2},{A})}\ket{\Psi}_7,\\\nonumber
\ket{\Psi}_8  & = &\alpha_{A}^{0}\alpha_{E}^{0}\ket{0}_A\ket{0}_E\ket{0}_1\ket{0}_2
+ \alpha_{A}^{0}\alpha_{E}^{1}\ket{2}_A\ket{1}_E\ket{3}_1\ket{2}_2
+ \alpha_{A}^{0}\alpha_{E}^{2}\ket{0}_A\ket{2}_E\ket{2}_1\ket{0}_2\\\nonumber
&+& \alpha_{A}^{0}\alpha_{E}^{3}\ket{2}_A\ket{3}_E\ket{1}_1\ket{2}_2
+ \alpha_{A}^{1}\alpha_{E}^{0}\ket{1}_A\ket{0}_E\ket{1}_1\ket{1}_2
+ \alpha_{A}^{1}\alpha_{E}^{1}\ket{3}_A\ket{1}_E\ket{0}_1\ket{3}_2\\\nonumber
&+&\alpha_{A}^{1}\alpha_{E}^{2}\ket{1}_A\ket{2}_E\ket{3}_1\ket{1}_2
+\alpha_{A}^{1}\alpha_{E}^{3}\ket{3}_A\ket{3}_E\ket{2}_1\ket{3}_2
+ \alpha_{A}^{2}\alpha_{E}^{0}\ket{2}_A\ket{0}_E\ket{2}_1\ket{2}_2\\\nonumber
&+& \alpha_{A}^{2}\alpha_{E}^{1}\ket{0}_A\ket{1}_E\ket{1}_1\ket{0}_2
+\alpha_{A}^{2}\alpha_{E}^{2}\ket{2}_A\ket{2}_E\ket{0}_1\ket{2}_2
+\alpha_{A}^{2}\alpha_{E}^{3}\ket{0}_A\ket{3}_E\ket{3}_1\ket{0}_2\\\nonumber
&+&\alpha_{A}^{3}\alpha_{E}^{0}\ket{3}_A\ket{0}_E\ket{3}_1\ket{3}_2
+ \alpha_{A}^{3}\alpha_{E}^{1}\ket{1}_A\ket{1}_E\ket{2}_1\ket{1}_2
+\alpha_{A}^{3}\alpha_{E}^{2}\ket{3}_A\ket{2}_E\ket{1}_1\ket{3}_2\\
&+&\alpha_{A}^{3}\alpha_{E}^{3}\ket{1}_A\ket{3}_E\ket{0}_1\ket{1}_2.
\end{eqnarray}
Finally, we apply a GXOR gate between $R_1$ and the agent,
\begin{eqnarray}
\label{eq29b}\nonumber
\ket{\Psi}_9  & = & U^{\rm{GXOR}}_{({R_1},{A})}\ket{\Psi}_8,\\\nonumber
\ket{\Psi}_9  & = &\alpha_{A}^{0}\alpha_{E}^{0}\ket{0}_A\ket{0}_E\ket{0}_1\ket{0}_2
+ \alpha_{A}^{0}\alpha_{E}^{1}\ket{1}_A\ket{1}_E\ket{3}_1\ket{2}_2
+ \alpha_{A}^{0}\alpha_{E}^{2}\ket{2}_A\ket{2}_E\ket{2}_1\ket{0}_2\\\nonumber
&+& \alpha_{A}^{0}\alpha_{E}^{3}\ket{3}_A\ket{3}_E\ket{1}_1\ket{2}_2
+ \alpha_{A}^{1}\alpha_{E}^{0}\ket{0}_A\ket{0}_E\ket{1}_1\ket{1}_2
+ \alpha_{A}^{1}\alpha_{E}^{1}\ket{1}_A\ket{1}_E\ket{0}_1\ket{3}_2\\\nonumber
&+&\alpha_{A}^{1}\alpha_{E}^{2}\ket{2}_A\ket{2}_E\ket{3}_1\ket{1}_2
+\alpha_{A}^{1}\alpha_{E}^{3}\ket{3}_A\ket{3}_E\ket{2}_1\ket{3}_2
+ \alpha_{A}^{2}\alpha_{E}^{0}\ket{0}_A\ket{0}_E\ket{2}_1\ket{2}_2\\\nonumber
&+& \alpha_{A}^{2}\alpha_{E}^{1}\ket{1}_A\ket{1}_E\ket{1}_1\ket{0}_2
+\alpha_{A}^{2}\alpha_{E}^{2}\ket{2}_A\ket{2}_E\ket{0}_1\ket{2}_2
+\alpha_{A}^{2}\alpha_{E}^{3}\ket{3}_A\ket{3}_E\ket{3}_1\ket{0}_2\\\nonumber
&+&\alpha_{A}^{3}\alpha_{E}^{0}\ket{0}_A\ket{0}_E\ket{3}_1\ket{3}_2
+ \alpha_{A}^{3}\alpha_{E}^{1}\ket{1}_A\ket{1}_E\ket{2}_1\ket{1}_2
+\alpha_{A}^{3}\alpha_{E}^{2}\ket{2}_A\ket{2}_E\ket{1}_1\ket{3}_2\\
&+&\alpha_{A}^{3}\alpha_{E}^{3}\ket{3}_A\ket{3}_E\ket{0}_1\ket{1}_2.
\end{eqnarray}
As we can see, based in the quantum protocol described previously (see \textbf{Fig \ref{Fig3}}), we have shown that for a fixed number of interactions, we obtain maximal learning fidelity even though the system has an arbitrary dimension.
\section*{Quantum reinforcement learning in multiqudit systems}
In the previous section, we proved that for an agent and environment composed of a multilevel system each, the quantum reinforcement learning protocol entails maximal learning fidelity for a fixed number of steps, irrespective of the dimension. Here, using this result, we also prove that for more than one multilevel system in agent, environment, and register subspaces, the number of steps is also fixed and scales with the number of individual subsystems that compose both agent and environment subsystems. To be more specific, in the single-multilevel case the needed total steps are nine. For two multilevel systems, we show that the number of required steps are eighteen, and in general, 9$n$, with $n$ being the number of multilevel subsystems. The possible initial states of our protocol consist in arbitrary superpositions for both agent and environment states and the register states are in their ground state,
\begin{eqnarray}
\label{eq37}
\ket{\Psi_0} & = & \sum_{n,m=0}^{N-1}\sum_{p,q=0}^{N-1}\alpha_{A}^{nm}\alpha_{E}^{pq}\ket{n}_A\ket{m}_A\ket{p}_E\ket{q}_E\ket{0}_1\ket{0}_2\ket{0}_3\ket{0}_4.
\end{eqnarray}
The first step in the protocol consists in encoding the environment information in the register states. This is done by applying a pair of GXOR gates. The gates are applied in the environment-register subspace, while the interaction in this case is the same as the one described previously. Namely, $E_1$ controls $R_1$ and $E_2$ controls $R_2$. 
\begin{eqnarray}
\label{eq38}\nonumber
\ket{\Psi_1} & = &  U^{\rm{GXOR}}_{(E_2,{R_2})} U^{\rm{GXOR}}_{(E_1,{R_1})}\ket{\Psi_0},\\
\ket{\Psi_1} & = &\sum_{n,m=0}^{N-1}\sum_{p,q=0}^{N-1}\alpha_{A}^{nm}\alpha_{E}^{pq}\ket{n}_A\ket{m}_A\ket{p}_E\ket{q}_E\ket{p}_1\ket{q}_2\ket{0}_3\ket{0}_4.
\end{eqnarray}
Similarly, in the second step we encode the environment information in the other two registers ($R_3$ and $R_4$) through GXOR gates. Here, the control system is the environment while the targets are the registers.
\begin{eqnarray}
\label{eq39}\nonumber
\ket{\Psi_2} & = &  U^{\rm{GXOR}}_{(E_2,{R_4})} U^{\rm{GXOR}}_{(E_1,{R_3})}\ket{\Psi_1},\\
\ket{\Psi_2} & = & \sum_{n,m=0}^{N-1}\sum_{p,q=0}^{N-1}\alpha_{A}^{nm}\alpha_{E}^{pq}\ket{n}_A\ket{m}_A\ket{p}_E\ket{q}_E\ket{p}_1\ket{q}_2\ket{p}_3\ket{q}_4.
\end{eqnarray}
Subsequently, a part of the register subspace is updated conditional on the agent information. Therefore, we apply a pair of GXOR gates on the agent-register subspace. In this case, agents $A_1$ and $A_2$ are controls and registers $R_1$ and $R_2$ targets.
\begin{eqnarray}
\label{eq40}\nonumber
\ket{\Psi_3} & = &  U^{\rm{GXOR}}_{(A_2,{R_2})} U^{\rm{GXOR}}_{(A_1,{R_1})}\ket{\Psi_2},\\
\ket{\Psi_3} & = & \sum_{n,m=0}^{N-1}\sum_{p,q=0}^{N-1}\alpha_{A}^{nm}\alpha_{E}^{pq}\ket{n}_A\ket{m}_A\ket{p}_E\ket{q}_E\ket{n\ominus p}_1\ket{m\ominus q}_2\ket{p}_3\ket{q}_4.
\end{eqnarray}
Now, we update the register subspace considering interactions between register components which have been acted upon with the same part of the environment. Namely, the register $R_3$ will be updated with the control of $R_1$ (Similarly with $R_4$ being controlled with $R_2$).
\begin{eqnarray}
\label{eq41}\nonumber
\ket{\Psi_4} & = &  U^{\rm{GXOR}}_{(R_2,{R_4})} U^{\rm{GXOR}}_{(R_1,{R_3})}\ket{\Psi_3},\\
\ket{\Psi_4} & = & \sum_{n,m=0}^{N-1}\sum_{p,q=0}^{N-1}\alpha_{A}^{nm}\alpha_{E}^{pq}\ket{n}_A\ket{m}_A\ket{p}_E\ket{q}_E\ket{n\ominus p}_1\ket{m\ominus q}_2\ket{n\ominus2p}_3\ket{m\ominus2q}_4.
\end{eqnarray}
Subsequently, we need to apply successive interactions between agent states and register states to obtain maximal learning fidelity. We show that applying the same interactions as for the single multilevel case for the triplet formed by agent $A_1$ with the environment parts $R_1$ and $R_3$ (similarly $A_2$ with $R_2$ and $R_4$), the maximal learning fidelity is reached. It is straightforward to show that
\begin{eqnarray}\nonumber
\ket{\Psi_9} &=&U^{\rm{GXOR}}_{(R_2,A_2)} U^{\rm{GXOR}}_{(R_1,A_1)}U^{\rm{GXOR}}_{(R_4,A_2)} U^{\rm{GXOR}}_{(R_3,A_1)}\times\\\nonumber
&&U^{\rm{GXOR}}_{(R_2,A_2)} U^{\rm{GXOR}}_{(R_1,A_1)}U^{\rm{GXOR}}_{(R_4,A_2)} U^{\rm{GXOR}}_{(R_3,A_1)}U^{\rm{GXOR}}_{(R_2,A_2)}\times\\\nonumber
&&U^{\rm{GXOR}}_{(R_1,A_1)}\ket{\Psi_4},\\\nonumber
\ket{\Psi_9} &=& \sum_{n,m=0}^{N-1}\sum_{p,q=0}^{N-1}\alpha_{A}^{nm}\alpha_{E}^{pq}\ket{p}_A\ket{q}_A\ket{p}_E\ket{q}_E\ket{n\ominus p}_1\ket{m\ominus q}_2\ket{n\ominus2p}_3\ket{m\ominus2q}_4\\.
\end{eqnarray}
Summarizing, for the case studied in this section, we demonstrate that the number of operations required to obtain maximal learning fidelity does not depend on the learning unit dimension and it is equal to eighteen operations, which correspond to the double of the required steps in the single multiqubit case. It is straightforward to realize that the number of needed operations to achieve maximal learning fidelity in a machine learning protocol composed by $n$ subsystems for agent and environment is equal to $9n$. Namely, the number of operations scales polynomially, indeed linearly, with the number of subsystems.
\section*{Quantum reinforcement learning in larger environments}
Up to now, the quantum reinforcement learning protocol described here always considers that the agent and the environment have the same number of subsystems, as well as the same dimension. In these cases, we have shown that by adding more system registers the quantum protocol improves in the sense that only one iteration and one measurement is enough to obtain maximal learning fidelity. Nevertheless, in more realistic scenarios, the agent must adapt to larger or more complex surroundings. Here, we discuss the situation where the environment has more subsystems than the agent, and therefore a larger dimension. As the environment has more information than the agent, it is expect that not all available surrounding information will be transferred to the agent. Indeed, we prove that by depending on the register-environment interaction, the agent can encode the information from one specific part of the environment. In this case, unlike the protocol previously discussed, we achieve maximal learning fidelity after applying one measurement and a rewarding iteration (feedback).

\begin{figure}[!h]
\centering
\includegraphics[width=1\linewidth]{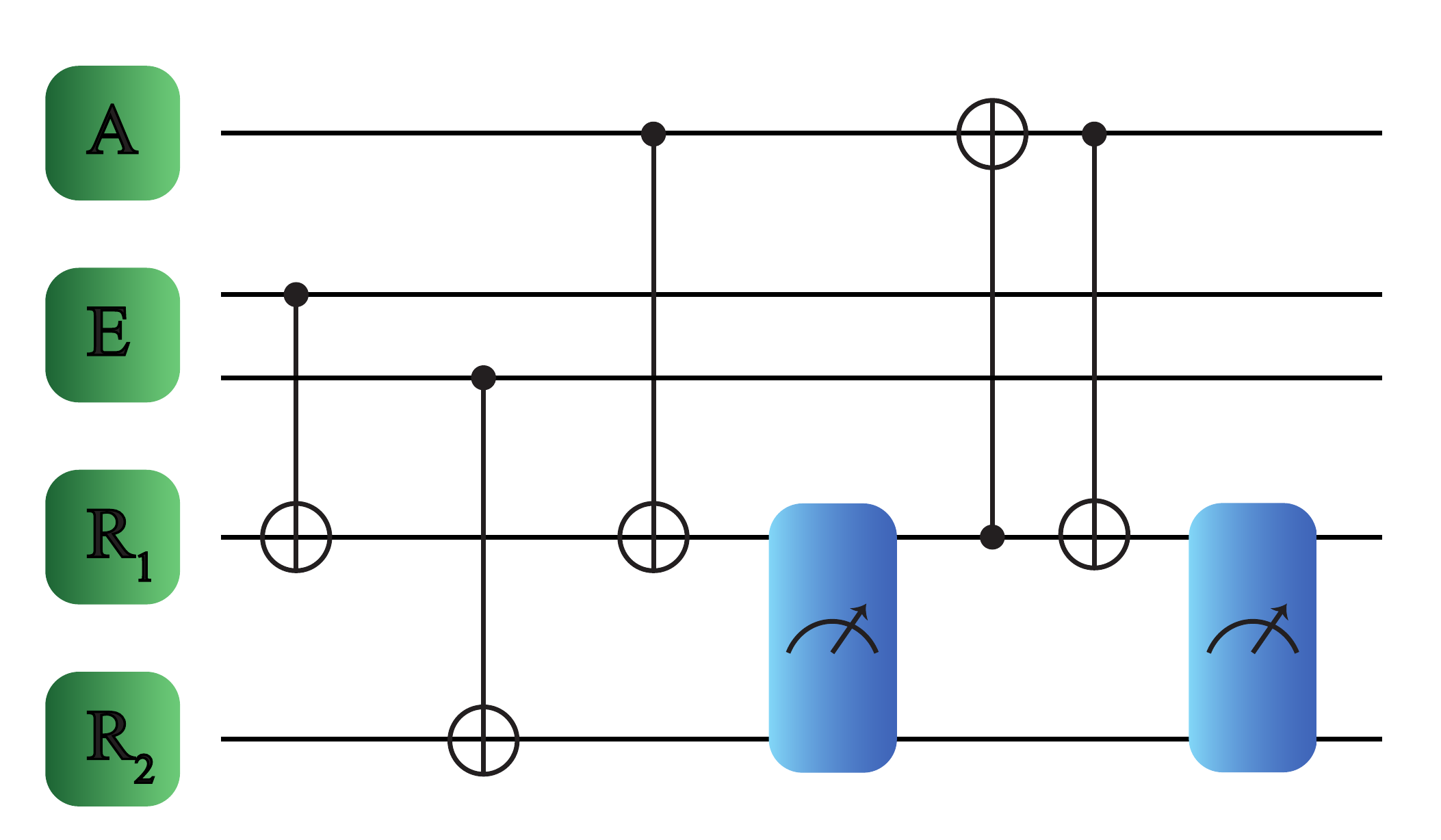}
\caption{\textbf{Quantum reinforcement learning for larger environment systems}. The systems involved are denoted as agent A, environment E and registers $R_1,R_2$, where E contains now two qubits while A just one. The logical gates applied between the different subsystems are CNOT gates. In this case, to obtain maximal learning fidelity, it is required to perform two separate measurements denoted by the blue boxes.}
\label{Fig4}
\end{figure}

The proposed quantum protocol is shown in \textbf{Fig. \ref{Fig4}}. Here, one two-level system forms the agent, while register and environment are constituted each by two qubits. Each environment qubit interacts with one qubit from the register, such that this interaction updates the registers conditionally to the environment information. Then, one part of the register subspace is also upgraded conditionally to the agent state. Subsequently, we perform a measurement on the register subspace, such that depending on the measurement outcomes we apply a conditional operation in the agent-register subspace until the agent adapts to a specific part of the environment. To illustrate this, let us introduce a possible agent-register-subspace state which has the following form,

\begin{eqnarray}
\label{eq46}
&&\ket{A} = \alpha_{A}^{0}\ket{0}_A + \alpha_{A}^1\ket{1}_A\\
&&\ket{E} = \alpha_{E}^{00}\ket{00}_E+\alpha_{E}^{01}\ket{01}_E+ \alpha_{E}^{10}\ket{10}_E+\alpha_{E}^{11}\ket{11}_E\\
&&\ket{R} = \ket{0}_1\ket{0}_2,\\
&&\ket{\Psi}_0=\ket{A}\ket{E}\ket{R}.
\end{eqnarray} 
The first step is to transfer quantum information from the environment onto the registers. This is done by applying a pair of CNOT gates in the environment-register subspaces,
\begin{eqnarray}
\label{eq47}\nonumber
\ket{\Psi}_1 & = & U^{\rm{CNOT}}_{({E},{R_2})}U^{\rm{CNOT}}_{({E},{R_1})}\ket{\Psi}_0,\\\nonumber
\ket{\Psi}_1 & = & (\alpha_{A}^{0}\ket{0}_A+\alpha_{A}^{1}\ket{1}_A)\\\nonumber
&&(\alpha_{E}^{00}\ket{00}_E\ket{0}_1\ket{0}_2+\alpha_{E}^{01}\ket{01}_E\ket{0}_1\ket{1}_2+ \alpha_{E}^{10}\ket{10}_E\ket{1}_1\ket{0}_2+\alpha_{E}^{11}\ket{11}_E\ket{1}_1\ket{1}_2).\\
\end{eqnarray}
Subsequently, the register $R_1$ is updated conditionally to the agent information. Therefore, a CNOT gate is applied in the agent-register subspace, where the agent qubit is the control and the register $R_1$ is the target,
\begin{eqnarray}
\label{eq48}\nonumber
\ket{\Psi}_2 & = & U^{\rm{CNOT}}_{({A},{R_1})}\ket{\Psi}_1,\\\nonumber
\ket{\Psi}_2 & = &\alpha_{A}^{0}\alpha_{E}^{00}\ket{0}_A\ket{00}_E\ket{0}_1\ket{0}_2+\alpha_{A}^{0}\alpha_{E}^{01}\ket{0}_A\ket{01}_E\ket{0}_1\ket{1}_2\\\nonumber
&+& \alpha_{A}^{0}\alpha_{E}^{10}\ket{0}_A\ket{10}_E\ket{1}_1\ket{0}_2+\alpha_{A}^{0}\alpha_{E}^{11}\ket{0}_A\ket{11}_E\ket{1}_1\ket{1}_2\\\nonumber
&+&\alpha_{A}^{1}\alpha_{E}^{00}\ket{1}_A\ket{00}_E\ket{1}_1\ket{0}_2+\alpha_{A}^{1}\alpha_{E}^{01}\ket{1}_A\ket{01}_E\ket{1}_1\ket{1}_2\\
&+& \alpha_{A}^{1}\alpha_{E}^{10}\ket{1}_A\ket{10}_E\ket{0}_1\ket{0}_2+\alpha_{A}^{1}\alpha_{E}^{11}\ket{1}_A\ket{11}_E\ket{0}_1\ket{1}_2.
\end{eqnarray}
Afterwards, we perform a measurement on the register subspace. In this case, the wave function is projected into the four possible measurement outcomes,
\begin{eqnarray}
\label{eq49}\nonumber
\rm{M}_1 &=& (\alpha_{A}^{0}\alpha_{E}^{00}\ket{0}_A\ket{00}_E + \alpha_{A}^{1}\alpha_{E}^{10}\ket{1}_A\ket{10}_E)\ket{0}_1\ket{0}_2\\\nonumber
&=& (\alpha_{A}^{0}\alpha_{E}^{00}\ket{0}_A\ket{0}_{E_1} + \alpha_{A}^{1}\alpha_{E}^{10}\ket{1}_A\ket{1}_{E_1})\ket{0}_{E_2}\ket{0}_1\ket{0}_2,\\\nonumber
\rm{M}_2 &=& (\alpha_{A}^{0}\alpha_{E}^{01}\ket{0}_A\ket{01}_E + \alpha_{A}^{1}\alpha_{E}^{11}\ket{1}_A\ket{11}_E)\ket{0}_1\ket{1}_2\\\nonumber
&=& (\alpha_{A}^{0}\alpha_{E}^{01}\ket{0}_A\ket{0}_{E_1} + \alpha_{A}^{1}\alpha_{E}^{11}\ket{1}_A\ket{1}_{E_1})\ket{1}_{E_2}\ket{0}_1\ket{1}_2,\\\nonumber
\rm{M}_3 &=& (\alpha_{A}^{1}\alpha_{E}^{00}\ket{1}_A\ket{00}_E + \alpha_{A}^{0}\alpha_{E}^{10}\ket{0}_A\ket{10}_E)\ket{1}_1\ket{0}_2\\\nonumber
&=& (\alpha_{A}^{1}\alpha_{E}^{00}\ket{1}_A\ket{0}_{E_1} + \alpha_{A}^{0}\alpha_{E}^{10}\ket{0}_A\ket{1}_{E_1})\ket{0}_{E_2}\ket{1}_1\ket{0}_2,\\\nonumber
\rm{M}_4 &=& (\alpha_{A}^{0}\alpha_{E}^{11}\ket{0}_A\ket{11}_E + \alpha_{A}^{1}\alpha_{E}^{01}\ket{1}_A\ket{01}_E)\ket{1}_1\ket{1}_2\\
&=& (\alpha_{A}^{0}\alpha_{E}^{11}\ket{0}_A\ket{1}_{E_1} + \alpha_{A}^{1}\alpha_{E}^{01}\ket{1}_A\ket{0}_{E_1})\ket{1}_{E_2}\ket{1}_1\ket{1}_2.
\end{eqnarray}
As we can see, the projective measurement on the register subspace produces that agent and one part of the environment subspace ($E_1$) is in an entangled state. At this stage, we can apply the rewarding criterion which consists in performing a CNOT gate operation in the register-agent subspace. The register qubit $R_1$ is the control and the agent is the target,
\begin{eqnarray}
\label{eq50}\nonumber
&&\rm{M}_{1a}=U^{\rm{CNOT}}_{({R_1},{A})}\rm{M}_1=(\alpha_{A}^{0}\alpha_{E}^{00}\ket{0}_A\ket{0}_{E_1} + \alpha_{A}^{1}\alpha_{E}^{10}\ket{1}_A\ket{1}_{E_1})\ket{0}_{E_2}\ket{0}_1\ket{0}_2,\\\nonumber
&&\rm{M}_{2a}=U^{\rm{CNOT}}_{({R_1},{A})}\rm{M}_2=(\alpha_{A}^{0}\alpha_{E}^{01}\ket{0}_A\ket{0}_{E_1} + \alpha_{A}^{1}\alpha_{E}^{11}\ket{1}_A\ket{1}_{E_1})\ket{1}_{E_2}\ket{0}_1\ket{1}_2,\\\nonumber
&&\rm{M}_{3a}=U^{\rm{CNOT}}_{({R_1},{A})}\rm{M}_3=(\alpha_{A}^{1}\alpha_{E}^{00}\ket{0}_A\ket{0}_{E_1} + \alpha_{A}^{0}\alpha_{E}^{10}\ket{1}_A\ket{1}_{E_1})\ket{0}_{E_2}\ket{1}_1\ket{0}_2,\\
&&\rm{M}_{4a}=U^{\rm{CNOT}}_{({R_1},{A})}\rm{M}_4=(\alpha_{A}^{0}\alpha_{E}^{11}\ket{1}_A\ket{1}_{E_1} + \alpha_{A}^{1}\alpha_{E}^{01}\ket{0}_A\ket{0}_{E_1})\ket{1}_{E_2}\ket{1}_1\ket{1}_2.
\end{eqnarray}
Finally, we perform a CNOT gate in the agent-register subspace to obtain orthogonal measurement outcomes. The qubit agent is the control and the qubit register $R_1$ is the target, according to
\begin{eqnarray}
\label{eq51}\nonumber
&&\rm{M}_{1b}=U^{\rm{CNOT}}_{({A},{R_1})}\rm{M}_{1a}=\alpha_{A}^{0}\alpha_{E}^{00}\ket{0}_A\ket{00}_E\ket{0}_1\ket{0}_2 + \alpha_{A}^{1}\alpha_{E}^{10}\ket{1}_A\ket{10}_E\ket{1}_1\ket{0}_2,\\\nonumber
&&\rm{M}_{2b}=U^{\rm{CNOT}}_{({A},{R_1})}\rm{M}_{2a}=\alpha_{A}^{0}\alpha_{E}^{01}\ket{0}_A\ket{01}_E\ket{0}_1\ket{1}_2  + \alpha_{A}^{1}\alpha_{E}^{11}\ket{1}_A\ket{11}_E\ket{1}_1\ket{1}_2,\\\nonumber
&&\rm{M}_{3b}=U^{\rm{CNOT}}_{({A},{R_1})}\rm{M}_{3a}=\alpha_{A}^{1}\alpha_{E}^{00}\ket{0}_A\ket{00}_E\ket{1}_1\ket{0}_2  + \alpha_{A}^{0}\alpha_{E}^{10}\ket{1}_A\ket{10}_E\ket{0}_1\ket{0}_2,\\
&&\rm{M}_{4b}=U^{\rm{CNOT}}_{({A},{R_1})}\rm{M}_{4a}=\alpha_{A}^{1}\alpha_{E}^{01}\ket{0}_A\ket{01}_E\ket{1}_1\ket{1}_2  + \alpha_{A}^{0}\alpha_{E}^{11}\ket{1}_A\ket{11}_E\ket{0}_1\ket{1}_2.
\end{eqnarray}
In this quantum reinforcement learning protocol, we perform interactions between the environment and the register subspaces. Nevertheless, the agent is updated only regarding the information encoded in register $R_1$. Thus, the maximal learning fidelity is achieved with respect to the first qubit of the environment.

Let us now consider another configuration similar to the one studied previously in this article, where the register is formed by a larger number of subsystems than the environment. Here, additionally, the environment we consider is larger than the agent. We prove that, for this system configuration, maximal learning fidelity between the agent and one part of the environment is achieved in one rewarding process. For this configuration, the maximal fidelity does not depend on the entanglement present in the agent-environment subspace. The general agent-register-environment state is
\begin{eqnarray}
&&\ket{A}=\alpha_{A}^{0}\ket{0}_A + \alpha_{A}^{1}\ket{1}_A,\\
&&\ket{E} =(\alpha_{E}^{0}\ket{0}_{E_1} + \alpha_{E}^{1}\ket{1}_{E_1})\ket{0}_{E_2} + (\beta_{E}^{0}\ket{0}_{E_1} + \beta_{E}^{1}\ket{1}_{E_1})\ket{1}_{E_2},\\
&&\ket{R} =\ket{0}_1\ket{0}_2\ket{0}_3\ket{0}_4,\\
&&\ket{\Psi}_0=\ket{A}\ket{E}\ket{R}.
\end{eqnarray}
The quantum protocol consists in updating the registers $R_{1,2}$ conditionally to the environment state $E_{1,2}$,
\begin{eqnarray}
\nonumber
\ket{\Psi}_1 & = & U^{\rm{CNOT}}_{({E_2},{R_2})}U^{\rm{CNOT}}_{({E_1},{R_1})}\ket{\Psi}_0,\\\nonumber
\ket{\Psi}_1 & = & (\alpha_{A}^{0}\ket{0}_A+\alpha_{A}^{1}\ket{1}_A)
(\alpha_{E}^{0}\ket{0}_{E_1}\ket{0}_{E_2}\ket{0}_1\ket{0}_2\ket{0}_3\ket{0}_4 + \alpha_{E}^{1}\ket{1}_{E_1}\ket{0}_{E_2}\ket{1}_1\ket{0}_2\ket{0}_3\ket{0}_4\\
&&+ \beta_{E}^{0}\ket{0}_{E_1}\ket{1}_{E_2}\ket{0}_1\ket{1}_2\ket{0}_3\ket{0}_4 + \beta_{E}^{1}\ket{1}_{E_1}\ket{1}_{E_2}\ket{1}_1\ket{1}_2\ket{0}_3\ket{0}_4).
\end{eqnarray}
After this, we also update the information of the registers $R_{3,4}$ conditionally to the environment state $E_{1,2}$,
\begin{eqnarray}
\nonumber
\ket{\Psi}_2 & = & U^{\rm{CNOT}}_{({E_2},{R_4})}U^{\rm{CNOT}}_{({E_1},{R_3})}\ket{\Psi}_1,\\\nonumber
\ket{\Psi}_2 & = & (\alpha_{A}^{0}\ket{0}_A+\alpha_{A}^{1}\ket{1}_A)
(\alpha_{E}^{0}\ket{0}_{E_1}\ket{0}_{E_2}\ket{0}_1\ket{0}_2\ket{0}_3\ket{0}_4 + \alpha_{E}^{1}\ket{1}_{E_1}\ket{0}_{E_2}\ket{1}_1\ket{0}_2\ket{1}_3\ket{0}_4\\
&&+ \beta_{E}^{0}\ket{0}_{E_1}\ket{1}_{E_2}\ket{0}_1\ket{1}_2\ket{0}_3\ket{1}_4 + \beta_{E}^{1}\ket{1}_{E_1}\ket{1}_{E_2}\ket{1}_1\ket{1}_2\ket{1}_3\ket{1}_4).
\end{eqnarray}
Now, the register $R_1$ is updated conditionally to the agent state,
\begin{eqnarray}
\label{step}\nonumber
\ket{\Psi}_3 & = & U^{\rm{CNOT}}_{({A},{R_1})}\ket{\Psi}_2,\\\nonumber
\ket{\Psi}_3 & = & \alpha_{A}^{0}\alpha_{E}^{0}\ket{0}_{A}\ket{0}_{E_1}\ket{0}_{E_2}\ket{0}_1\ket{0}_2\ket{0}_3\ket{0}_4 + \alpha_{A}^{0}\alpha_{E}^{1}\ket{0}_{A}\ket{1}_{E_1}\ket{0}_{E_2}\ket{1}_1\ket{0}_2\ket{1}_3\ket{0}_4\\\nonumber
&+&\alpha_{A}^{0}\beta_{E}^{0}\ket{0}_{A}\ket{0}_{E_1}\ket{1}_{E_2}\ket{0}_1\ket{1}_2\ket{0}_3\ket{1}_4+
\alpha_{A}^{0}\beta_{E}^{1}\ket{0}_{A}\ket{1}_{E_1}\ket{1}_{E_2}\ket{1}_1\ket{1}_2\ket{1}_3\ket{1}_4\\\nonumber
&+&\alpha_{A}^{1}\alpha_{E}^{0}\ket{1}_{A}\ket{0}_{E_1}\ket{0}_{E_2}\ket{1}_1\ket{0}_2\ket{0}_3\ket{0}_4 + 
\alpha_{A}^{1}\alpha_{E}^{1}\ket{1}_{A}\ket{1}_{E_1}\ket{0}_{E_2}\ket{0}_1\ket{0}_2\ket{1}_3\ket{0}_4 \\\nonumber
&+&\alpha_{A}^{1}\beta_{E}^{0}\ket{1}_{A}\ket{0}_{E_1}\ket{1}_{E_2}\ket{1}_1\ket{1}_2\ket{0}_3\ket{1}_4 +
\alpha_{A}^{1}\beta_{E}^{1}\ket{1}_{A}\ket{1}_{E_1}\ket{1}_{E_2}\ket{0}_1\ket{1}_2\ket{1}_3\ket{1}_4.\\
\end{eqnarray}
Then, the next step would consist in updating a part of the register subspace from the information encoded in the other part. However, this step is not necessary because the number of terms in Eq.~(\ref{step}) is smaller than all the possible measurement outcomes in the register subspace. Thus, the register is always projected onto orthogonal measurement outcomes. On the other hand, we update the agent state from the information encoding in the register $R_1$. Therefore, we perform a CNOT gate in the register-agent subspace, where the register $R_1$ is the control and the agent is the target,
\begin{eqnarray}
\nonumber
\ket{\Psi}_4 & = & U^{\rm{CNOT}}_{({R_1},{A})}\ket{\Psi}_3,\\\nonumber
\ket{\Psi}_4 & = & \alpha_{A}^{0}\alpha_{E}^{0}\ket{0}_{A}\ket{0}_{E_1}\ket{0}_{E_2}\ket{0}_1\ket{0}_2\ket{0}_3\ket{0}_4 + \alpha_{A}^{0}\alpha_{E}^{1}\ket{1}_{A}\ket{1}_{E_1}\ket{0}_{E_2}\ket{1}_1\ket{0}_2\ket{1}_3\ket{0}_4\\\nonumber
&+& \alpha_{A}^{0}\beta_{E}^{0}\ket{0}_{A}\ket{0}_{E_1}\ket{1}_{E_2}\ket{0}_1\ket{1}_2\ket{0}_3\ket{1}_4 +\alpha_{A}^{0}\beta_{E}^{1}\ket{1}_{A}\ket{1}_{E_1}\ket{1}_{E_2}\ket{1}_1\ket{1}_2\ket{1}_3\ket{1}_4\\\nonumber
&+&\alpha_{A}^{1}\alpha_{E}^{0}\ket{0}_{A}\ket{0}_{E_1}\ket{0}_{E_2}\ket{1}_1\ket{0}_2\ket{0}_3\ket{0}_4
+\alpha_{A}^{1}\alpha_{E}^{1}\ket{1}_{A}\ket{1}_{E_1}\ket{0}_{E_2}\ket{0}_1\ket{0}_2\ket{1}_3\ket{0}_4 \\\nonumber
&+&\alpha_{A}^{1}\beta_{E}^{0}\ket{0}_{A}\ket{0}_{E_1}\ket{1}_{E_2}\ket{1}_1\ket{1}_2\ket{0}_3\ket{1}_4
+\alpha_{A}^{1}\beta_{E}^{1}\ket{1}_{A}\ket{1}_{E_1}\ket{1}_{E_2}\ket{0}_1\ket{1}_2\ket{1}_3\ket{1}_4.\\
\end{eqnarray}
By measuring the register subspace, we obtain that agent and environment qubit $E_1$ achieve maximal fidelity.
\section*{Quantum reinforcement learning for mixed states}
Let us consider now the situation where the environment evolves under a noisy mechanism (for qubit states, noisy mechanisms can be depolarizing noise as well as amplitude damping). In this case, the density matrix describing the environment state reads
\begin{eqnarray}
\label{eq53}
\rho=\begin{pmatrix}
\rho_{00} & \rho_{01} \\ 
\rho_{01}^{*} & \rho_{11}
\end{pmatrix} .
\end{eqnarray}
We focus now our attention in the application of the quantum reinforcement learning protocol in this type of state. We will show that, by adding more registers, two main results will be obtained. Firstly, even though the environment is in a mixed state, the learning fidelity will be maximal for any measurement outcome in the register basis. Additionally, the measurement outcomes provide relevant information about the coherences of the mixed state. To apply the quantum protocol, we express the mixed state in term of its (non-unique) purification, such as
\begin{eqnarray}
\label{eq6}
\ket{\Psi_{E+e}}&=&\bigg[\sqrt{\rho_{00}}\ket{0}_{E}+\frac{\rho_{10}}{\sqrt{\rho_{00}}}\ket{1}_{E}\bigg]\ket{e_1} + \bigg[\sqrt{\rho_{11}-\frac{|\rho_{10}|^2}{\rho_{00}}}\bigg]\ket{1}_{E}\ket{e_2},\\\nonumber
\ket{\psi_e} &=& \frac{\rho_{10}}{\sqrt{\rho_{00}}}\ket{e_1} + \bigg[\sqrt{\rho_{11}-\frac{|\rho_{10}|^2}{\rho_{00}}}\bigg]\ket{e_2}\rightarrow \ket{\Psi_{E+e}}= \sqrt{\rho_{00}}\ket{0}_{E}\ket{e_1} + \sqrt{\rho_{11}}\ket{1}_{E}\ket{\bar{\psi_e}}.\\
\end{eqnarray}
Here, $\ket{\bar{\psi_e}}$ is a normalized vector in the purification Hilbert space. As we can see, the coefficient of the quantum state written in its extended Hilbert space (environment + purification) depends only on the diagonal terms of the mixed state. Moreover, to obtain additional information about the mixed state, we need to perform unitary transformations on it in such a way that the information related to the coherences is in the diagonal of the state after the transformation. To be more specific, we need to perform unitary transformations such that the mixed state can be written as follows,
\begin{eqnarray}
\label{eq5}
\bar{\rho}\rightarrow U_{y}\rho U_{y}^{\dag}=\frac{1}{2}\begin{pmatrix}
1+(\rho_{01}+\rho_{01}^{*}) & \rho_{11}-\rho_{00} +(\rho_{01}-\rho_{01}^{*}) \\ 
\rho_{11}-\rho_{00} -(\rho_{01}-\rho_{01}^{*}) & 1-(\rho_{01}+\rho_{01}^{*})
\end{pmatrix} ,
\end{eqnarray}
\begin{eqnarray}
\label{eq5}
\tilde{\rho}\rightarrow U_{x}\rho U_{x}^{\dag}=\frac{1}{2}\begin{pmatrix}
1-i(\rho_{01}-\rho_{01}^{*}) & \rho_{01}+\rho_{01}^{*} +i(\rho_{11}-\rho_{00}) \\ 
\rho_{01}+\rho_{01}^{*} -i(\rho_{11}-\rho_{00}) & 1+i(\rho_{01}-\rho_{01}^{*})
\end{pmatrix} .
\end{eqnarray}
To carry out this task, we need to add three more registers, where each of them has the function to encode information of diagonal, real, and imaginary part of the coherence terms, respectively. A possible state for the space composed of agent, mixed environment and register is given by
\begin{eqnarray}
\label{eq46}
&&\ket{A} = \alpha_{A}^{0}\ket{0}_A + \alpha_{A}^1\ket{1}_A,\\
&&\ket{\Psi_{E+e}}=\sqrt{\rho_{00}}\ket{0}_{E}\ket{e_1} + \sqrt{\rho_{11}}\ket{1}_{E}\ket{\bar{\psi_e}}\\
&&\ket{R} = \ket{0}_1\ket{0}_2\frac{1}{\sqrt{3}}(\ket{1}_3\ket{0}_4\ket{0}_5 + \ket{0}_3\ket{1}_4\ket{0}_5 + \ket{0}_3\ket{0}_4\ket{1}_5)\\
&&\ket{\Psi}_0=\ket{A}\ket{\Psi_{E+e}}\ket{R}.
\end{eqnarray} 
The first step is to apply a unitary transformation, which is conditional to the state of the register $R_3$, $R_4$ and $R_5$. In case that the register state is $\ket{1}_3\ket{0}_4\ket{0}_5$, we apply the transformation $\mathcal{U}_{1}=\mathbb{I}_{R_3}\otimes\mathbb{I}_{R_4}\otimes\mathbb{I}_{R_5}$. If the register state is in the state $\ket{0}_3\ket{1}_4\ket{0}_5$, we apply the transformation $\mathcal{U}_{2}=\mathbb{I}_{R_3}\otimes U_{y}\otimes\mathbb{I}_{R_5}$. Finally, if the register state is in the state $\ket{0}_3\ket{0}_4\ket{1}_5$ the unitary transformation is given by $\mathcal{U}_{3}=\mathbb{I}_{R_3}\otimes\mathbb{I}_{R_4}\otimes U_{x}$. Hence, the state after this transformation is given by unitary transformation in the environment state according to
\begin{eqnarray}\nonumber
\ket{\Psi}_{1} & = & \ket{A}\ket{\psi_{E+e}}\ket{0}_1\ket{0}_2\ket{1}_3\ket{0}_4\ket{0}_5 + \ket{A}U_{y}\ket{\psi_{E+e}}\ket{0}_1\ket{0}_2\ket{0}_3\ket{1}_4\ket{0}_5\\\nonumber
&+& \ket{A}U_{x}\ket{\psi_{E+e}}\ket{0}_1\ket{0}_2\ket{0}_3\ket{0}_4\ket{1}_5,\\\nonumber
\ket{\Psi_1}  & = & \frac{1}{\sqrt{3}}(\alpha_{A}^{0}\ket{0}_A + \alpha_{A}^1\ket{1}_A)\big[(\sqrt{\rho_{00}}\ket{0}_{E}\ket{e_1} + \sqrt{\rho_{11}}\ket{1}_{E}\ket{\bar{\psi_e}})\ket{0}_1\ket{0}_2\ket{1}_3\ket{0}_4\ket{0}_5\\\nonumber
&&+\bigg(\sqrt{\frac{1}{2}+\operatorname{Re}(\rho_{01})}\ket{0}_{E}\ket{e_1} + \sqrt{\frac{1}{2}-\operatorname{Re}(\rho_{01})}\ket{1}_{E}\ket{\bar{\psi_e}}\bigg)\ket{0}_1\ket{0}_2\ket{0}_3\ket{1}_4\ket{0}_5\\\nonumber
&&+\bigg(\sqrt{\frac{1}{2}+\operatorname{Im}(\rho_{01})}\ket{0}_{E}\ket{e_1} + \sqrt{\frac{1}{2}-\operatorname{Im}(\rho_{01})}\ket{1}_{E}\ket{\bar{\psi_e}}\bigg)\ket{0}_1\ket{0}_2\ket{0}_3\ket{0}_4\ket{1}_5].\\
\end{eqnarray}
Afterwards, we apply the quantum protocol as we did in the first section. Namely, we first update the register conditionally to the information of the environment. Then, we update the register $R_1$ conditionally to the information of the agent. Subsequently, to obtain orthogonal measurement outcomes we perform CNOT gates in the register subspace ($R_1$ is the control and $R_2$ is the agent). Finally, the agent is updated in terms of the information encoded in register $R_1$ (where A is the target and $R_1$ is the control),
\begin{eqnarray}
\label{eqn15}\nonumber
\ket{\Psi}_5  & = & \frac{1}{\sqrt{3}}\bigg(\alpha_{A}^{0}\sqrt{\rho_{00}}\ket{0}_A\ket{0}_E\ket{e_1}\ket{0}_1\ket{0}_2\ket{1}_3\ket{0}_4\ket{0}_5\\\nonumber
&+&\alpha_{A}^{0}\sqrt{\rho_{11}}\ket{1}_A\ket{1}_E\ket{\bar{\psi_e}}\ket{1}_1\ket{0}_2\ket{1}_3\ket{0}_4\ket{0}_5\\\nonumber 
&+& \alpha_{A}^{1}\sqrt{\rho_{00}}\ket{0}_A\ket{0}_E\ket{e_1}\ket{1}_1\ket{1}_2\ket{1}_3\ket{0}_4\ket{0}_5 \\\nonumber
&+&\alpha_{A}^{1}\sqrt{\rho_{11}}\ket{1}_A\ket{1}_E\ket{\bar{\psi_e}}\ket{0}_1\ket{1}_2\ket{1}_3\ket{0}_4\ket{0}_5\\\nonumber
&+&\alpha_{A}^{0}{\sqrt{\frac{1}{2}+\operatorname{Re}(\rho_{01})}}\ket{0}_A\ket{0}_E\ket{e_1}\ket{0}_1\ket{0}_2\ket{0}_3\ket{1}_4\ket{0}_5\\\nonumber
&+& \alpha_{A}^{0}{\sqrt{\frac{1}{2}-\operatorname{Re}(\rho_{01})}}\ket{1}_A\ket{1}_E\ket{\bar{\psi_e}}\ket{1}_1\ket{0}_2\ket{0}_3\ket{1}_4\ket{0}_5\\\nonumber 
&+& \alpha_{A}^{1}{\sqrt{\frac{1}{2}+\operatorname{Re}(\rho_{01})}}\ket{0}_A\ket{0}_E\ket{e_1}\ket{1}_1\ket{1}_2\ket{0}_3\ket{1}_4\ket{0}_5 \\\nonumber
&+&\alpha_{A}^{1}{\sqrt{\frac{1}{2}-\operatorname{Re}(\rho_{01})}}\ket{1}_A\ket{1}_E\ket{\bar{\psi_e}}\ket{0}_1\ket{1}_2\ket{0}_3\ket{1}_4\ket{0}_5\\\nonumber
&+&\alpha_{A}^{0}{\sqrt{\frac{1}{2}+\operatorname{Im}(\rho_{01})}}\ket{0}_A\ket{0}_E\ket{e_1}\ket{0}_1\ket{0}_2\ket{0}_3\ket{0}_4\ket{1}_5\\\nonumber
&+&\alpha_{A}^{0}{\sqrt{\frac{1}{2}-\operatorname{Im}(\rho_{01})}}\ket{1}_A\ket{1}_E\ket{\bar{\psi_e}}\ket{1}_1\ket{0}_2\ket{0}_3\ket{0}_4\ket{1}_5\\\nonumber 
&+& \alpha_{A}^{1}{\sqrt{\frac{1}{2}+\operatorname{Im}(\rho_{01})}}\ket{0}_A\ket{0}_E\ket{e_1}\ket{1}_1\ket{1}_2\ket{0}_3\ket{0}_4\ket{1}_5 \\
&+& \alpha_{A}^{1}{\sqrt{\frac{1}{2}-\operatorname{Im}(\rho_{01})}}\ket{1}_A\ket{1}_E\ket{\bar{\psi_e}}\ket{0}_1\ket{1}_2\ket{0}_3\ket{0}_4\ket{1}_5\bigg).
\end{eqnarray}

This quantum reinforcement learning protocol exhibits two features. First, by performing projective measurements on registers $R_1$, $R_2$ and $R_3$, we recover the result studied in the first section, i.e., the learning fidelity is maximal independently of the measurement outcomes in the register subspace. The second feature is that, for specific measurement outcomes in a part of the register subspace, we obtain information about the population (diagonal) and the coherences (off-diagonal) of the mixed state. This feature can be used in problems such as partial cloning in cases where the system in which we can extract information evolves under loss mechanisms.

\section*{Analysis of implementation in quantum technologies}
An interesting result obtained in this manuscript is that in most of the cases, for the considered quantum reinforcement learning protocols, adding more registers improves the rewarding process. That is, via a purely unitary evolution, without coherent feedback, a maximally positively-correlated agent environment state is achieved, in the sense that the final agent contains the same quantum information as the considered final environment. This means that the agent has acquired the needed information about the environment and accordingly modified it, being this a quantum process. In our formalism, typically, one measurement at the end of the protocol is enough to obtain maximal learning fidelity in one iteration of the process. In this sense, several quantum architectures could benefit of this fact, given that coherent feedback is not needed in this case. For instance, we focus our attention in two prominent platforms, namely, trapped ions and superconducting circuits. 
\subsection*{Trapped ions}
As we have pointed out along the manuscript, the performance of our proposed quantum protocols is based on the quality of the quantum gates between different subsystems. In this case, the realization of high-fidelity quantum gates is essential to perform the quantum protocol proposed here. Technological progress in trapped ions has enabled to implement single ~\cite{PhysRevA.84.030303} and two-qubit quantum gates \cite{Benhelm2008} with a large fidelity. For the single-qubit gate, e.g., a Beryllium hyperfine transition can be driven with microwave fields or lasers, being the error associated with single-qubit gates below $10^{-4}$. For two-qubit gates, the use of either microwaves or a laser beam with modulated amplitude allows for the interaction of both qubits (electronic levels of, e.g., Beryllium or Calcium ions) at the same time. Adiabatic elimination of the motion allows one to obtain maximally entangled states of both ions. The fidelity of trapped-ion two-qubit gates can reach nowadays above 99.9{\%} \cite{PhysRevLett.117.060505,PhysRevLett.117.140501}. Trapped-ion technologies offer long coherences times, which can reach up to the range of seconds \cite{PhysRevLett.95.060502} for Calcium atoms. In addition, this platform enables state preparation and readout with high fidelity \cite{PhysRevLett.100.200502,Noek13,PhysRevLett.113.220501}. Here, the use of hyperfine states and the microwave fields improve the optical pumping fidelity and improve the relaxation time $T_1$ allowing to obtain fidelity readouts of 99.9999{\%} \cite{PhysRevLett.100.200502}.
\subsection*{Superconducting circuits}
As in trapped ions, the technological progress in superconducting circuits has grown significantly in the latter years. For instance, artificial atoms whose coherence times are in the microsecond range have been built in coplanar \cite{PhysRevLett.111.080502} and 3D architectures \cite{PhysRevLett.107.240501}. On the other hand, integrated Josephson quantum processors allows one to implement quantum gates between two-level systems even in cases where the qubits do not have identical frequencies, as well as making them interact via a quantum bus \cite{PhysRevA.69.062320}. The Xmon qubits achieve two-qubit gate fidelities above 99{\%}\cite{Barends2014, Barends2016}. These technological progresses have developed feedback loop control in this platform. This feedback protocol relies on high fidelity readout, as well as on conditional control on the outcome of a quantum non-demolition measurement \cite{PhysRevLett.109.240502,riste2015digital}. Even though in the quantum reinforcement learning protocols in this paper coherent feedback is not required, this may be a useful ingredient in other quantum reinforcement learning proposals~\cite{Lucas2017}.
\section*{Discussion}
In summary, we propose a protocol to perform quantum reinforcement learning which does not require coherent feedback and, therefore, may be implemented in a variety of quantum technologies. Our learning protocol, being mostly unitary (except with the final register measurement) considers learning in a loose sense: while it does not depend on feedback, the protocol achieves its aim regardless of the initial state of agent and environment. In this aspect, it is general, and obtains a similar goal than Ref.~\cite{Lucas2017} without the need of feedback, enabling its implementation in a variety of quantum platforms. We also point out that one may employ different performance measures than the one considered here, depending on the agent possible aims. Adding more registers than in previous proposals in the literature~\cite{Lucas2017}, the rewarding criterion can be applied at the end of the protocol, while agent and environment need not be measured directly, although only via the registers. We also obtain that when the considered systems are composed of qudits, the number of steps needed to obtain maximal learning fidelity is fixed in each qudit dimension and scales polynomially with the number of qudit subsystems. We consider as well environment states which are mixtures, while the agent can also in this case acquire the appropriate information from them. Theoretically, all the cases considered of qubit, multiqubit, qudit, and multiqudit, have many similarities. Even though the protocols are not directly transformable into one another, a $d$-dimensional qudit can be rewritten as a $\log_2 (d)$ multiqubit system, while a multiqudit system with $n$ qudits is equivalent to an $n \log_2 (d)$ multiqubit system. Therefore, in this respect, it is intuitive that the results for all these protocols (namely, that maximal fidelity can be attained) should be related. Nevertheless, it is valuable to show that the protocol can be scaled up to multiqudit systems with many parties and high dimensions, given that this will be an ultimate goal of a scalable quantum device.
 Implementations of these protocols in trapped ions and superconducting circuits seem feasible with current platforms. 


\section*{Author contributions}
F. A. C.-L. developed the protocol under the guidance of J. C. R., based on original ideas and feedback by L. L. All authors contributed to the development of the ideas, performance of the protocol and writing of the manuscript. E. S. supervised the project in all stages.
\nolinenumbers

\section*{Competing interests}

L. L. declares his affiliation with PLOS ONE as Academic Editor.

%
%
%

\end{document}